\documentclass{emulateapj}

\usepackage{aas_macros}
\usepackage{epstopdf}
\usepackage{epsfig}
\usepackage{natbib}
\usepackage{float}
\usepackage{gensymb}
\usepackage{amsmath}
\usepackage{lipsum}
\usepackage{times}
\usepackage{comment}
\usepackage{color}
\usepackage{multirow}
\usepackage{soul}
\usepackage{CJKutf8}            
\usepackage{xspace}

\usepackage{lineno}

\definecolor{purple}{RGB}{160,32,240}

\definecolor{purple2}{RGB}{120,72,240}

\begin{document}

\title{The Anisotropic Circumgalactic Medium of Sub-L$^*$ Galaxies}

\author{Huanian Zhang \begin{CJK*}{UTF8}{gkai} (张华年)\altaffilmark{1,3}
\end{CJK*}}

\author{Miao Li \begin{CJK*}{UTF8}{gkai} (李邈)\altaffilmark{2}
\end{CJK*}}

\author{Dennis Zaritsky\altaffilmark{3}}
\altaffiltext{1}{Department of Astronomy, 
Huazhong University of Science and Technology, Wuhan, Hubei 430074, China; huanian@hust.edu.cn}
\altaffiltext{2}{Insitute for Astronomy, School of Physics, 
Zhejiang University, Hangzhou, Zhejiang 310027, China; miaoli@zju.edu.cn}
\altaffiltext{3}{Steward Observatory, University of Arizona, Tucson, AZ 85719, USA; dennis.zaritsky@gmail.com}

\begin{abstract}
Using stacked emission line flux measurements of cool circumgalactic gas (CGM) in lower-mass galaxies ($10^{9.0} \le M_*/M_\odot \le 10^{10.2} $), we measure the dependence of the emission characteristics on orientation relative to the disk plane as a function of radius and compare to that we found previously for massive ($M_* > 10^{10.4} M_\odot$) early-type galaxies.  Although the line ratios (the lower [N {\small II}]/H$\alpha$ and [O {\small III}]/H$\beta$) suggest an overall softer ionizing source than in the more massive galaxies, consistent with previous findings, we find the same ionization hardening signature (a higher [N {\small II}]/H$\alpha$ ratio in the inner polar region) along the polar direction at small radii that we found for the more massive galaxies. The line ratio in the inner polar bin is distinct from that measured for the inner planar bin with 99.99\% confidence and with $>$ 99.9\% confidence we conclude that it lies outside the star formation regime.  The effective hardening of the ionization of the CGM along the polar axis, at small radii, could either indicate relic effects of AGN activity or shock ionization. In either case, this signature appears to be ubiquitous across the stellar mass range we are able to explore with our spectral stacking technique and currently available archival data. 
\end{abstract}


\section{Introduction}

Observational and theoretical studies of galaxies all indicate that galaxies contain large reservoirs of gas within their virial radii. This gas is referred to as the circumgalactic medium \citep[CGM; see][for a review]{CGM2017, CGM2023} and
it is critical in the evolution of  galaxies \citep[cf.][]{Donahue2022} because it provides the fuel for star formation.
Recent observations and simulations show that the CGM is a multi-scale, multi-phase reservoir of gas with complex dynamics. Observational constraints are scant and insufficient to fully constrain the range of physical processes envisioned to be relevant.

While inflowing gas from the CGM can
fuel star formation in the central galaxy, physical processes within that central galaxy can act to retard that process. For galaxies with stellar mass greater than $10^{10.4}$ M$_\odot$, there is active debate on the physical mechanism. While the dominant factor is thought to be the energy and momentum released by accretion of matter \citep{Morganti2017} onto a central supermassive black hole \citep[SMBH,][]{Fabian2012}, the influence of star formation activities \citep[stellar feedback, eg.,][]{Ceverino2009, Hopkins2012, Shen2013, Somerville2015, Naab2017} should not be ignored. For lower mass galaxies, the principal influence is thought to come from supernovae and stellar winds \citep{Li2017, Kim2018, Hu2019, Limiao2020, Li2020, Fielding2020}. 
Given the different geometries involved, central source vs. distributed sources, one might speculate that any resulting signatures of these processes in the CGM would be distinct in high spatial resolution line ratio maps of the CGM of high and low mass galaxies, particularly at smaller radii where such sources of ionization would dominate over the contribution from the extragalactic UV background.

Cosmological simulations and observations find that the  CGM is anisotropic.  Anisotropies are predicted in the metallicity \citep{Peroux2020}, density \citep{Nelson2019},  temperature \citep{Truong2021} and magnetic field \citep{Ramesh2023a} from cosmological simulations. Observations are beginning to measures, or at least infer, anisotropies of the CGM distribution \citep{Bordoloi2011, Ho2017,Martin2019, Pointon2019, Schroetter2019, Ignacio2021}. 
Of particular relevance to this study, we previously measured anisotropies in the emission line ratios of the CGM in massive early type galaxies
\citep{Zhang2022}.

Although the majority of CGM studies to date rely on measurements of ultraviolet (UV) absorption lines in the spectra of bright background objects \citep[e.g.][]{steidel2010,Bordoloi2011,zhu2013a,zhu2013b,Werk2013,Johnson2014,Werk2014,Johnson2015,werk16,croft2016,prochaska2017,Cai2017,Johnson2017,Chen2017a,lan2018,joshi2018,Chen2019,Zahedy2019,Dutta2020,CGMsquare2021,Norris2021,Qu2022}, the scarcity of suitable lines of sight has motivated a growing set of complementary studies focusing on measurements of emission lines. Optical emission lines originating in the CGM provide an opportunity to image the cool ($T \sim 10^4$ K) phase of the CGM across each galaxy, but are challenging to measure \citep{zhang2016}. 
In the local universe, H$\alpha$ emission has been detected in the CGM of {\sl individual} galaxies only when the systems are extreme, such as  in the NGC 4631/4656 group \citep{Donahue1995}, the starburst/merger NGC 6240 \citep{Yoshida2016}, the nearby edge-on galaxy UGC 7321 \citep{Fumagalli2017}, and a low-mass (M$_* \sim$ $6\times 10^6$ M$_\odot$) blue compact dwarf galaxy \citep{Herenz2023}. [O{ \small II}], H$\beta$ and [O{ \small III}] have also been detected for a star-burst sub-$L^*$ galaxy \citep{Nielsen2023}.  \cite{zhang2016} presented the first detection of  H$\alpha$ and [N{ \small II}] $\lambda$6583 emissions, from low redshift, normal galaxies extending out to a projected radius of $\sim$ 100 kpc by stacking a sample of millions of sightlines from the SDSS DR12 \citep{SDSS12}. 

Stacking archival spectra enables studies that address a range of topics, such as the ionization mechanism, the effect of  environment, the rate of gas inflow, and the effect of feedback \citep[][hereafter, Papers I, II, III, IV, V, VI, VII, VIII]{zhang2016,zhang2018a,Zhang2018b, Zhang2019, Zhang2020a, Zhang2020b, Zhang2021, Zhang2022}. Here we will attempt to measure the azimuthal distribution of the cool CGM around lower mass galaxies and compare to our previous results on high mass elliptical galaxies (Paper VIII).
As in most of those studies, we restrict the stacks to projected radii between 10 kpc, or $\sim 0.05R_{\rm vir}$ for $\sim L^*$ galaxies, and 50 kpc, or $\sim 0.25R_{\rm vir}$ for $\sim L^*$ galaxies. The inner boundary is set to mitigate contamination from the central galaxy and the outer boundary to mitigate contamination from from nearby galaxies (see Paper II for more discussion regarding the outer boundary). 

From among our previous studies, the most relevant ones here are Paper III and VIII.  Paper III presents a study of the physical properties of the CGM based on diagnostic line ratios and Paper VIII presents the anisotropic distribution of the CGM of massive galaxies and discusses the possible connection to the central supermassive black hole. Emission line ratios, like those used in the BPT diagram \citep{bpt}, provide guidance on the ionizing source of the halo gas. For example, the BPT diagram is used to distinguish between the two expected dominant sources of ionization, star formation and active galactic nuclei (AGN) \citep[{e.g.}][]{vo,kewley,kauffmann_agn}, in galaxies showing emission line spectra. 
In Paper III, we  found that lower mass galaxies, $M_* < 10^{10.4}$ M$_\odot$, have a CGM that is ionized by softer sources, similar to that found in star forming regions (H {\small II} regions), while higher mass galaxies, $M_* > 10^{10.4}$ M$_\odot$, have a CGM that is ionized by harder sources, similar to that found in AGN-hosting galaxies or in shocked regions. In Paper VIII, we examined the lines ratio as a function of projected radius and the orientation with respect to the major axis of the central galaxy and we found that diagnostic line ratios show stronger AGN ionization signatures along the polar direction at small radii than at other angles or radii with high confidence (99\%). Here we will utilize  the same diagnostic tool to investigate the signature of the line ratios for lower-mass galaxies.

This paper is organized as follows. In \S\ref{sec:dataAna} we present the data analysis, including sample selection and reprise the basics of our technique. In \S\ref{sec:results} we present our measurements and identify any statistically significant differences as a function of azimuthal angle and radius. In \S\ref{sec:simu} we discuss implications in the context of simulations and observations. In \S\ref{sec:sum}, we summarize and conclude.
Throughout this paper, we adopt a $\Lambda$CDM cosmology with parameters
$\Omega_m$ = 0.3, $\Omega_\Lambda =$ 0.7, $\Omega_k$ = 0 and the dimensionless Hubble constant $h = $ 0.7 \citep[cf.][]{riess,Planck2018}.

\section{Data Analysis}
\label{sec:dataAna}

We continue as we have in Papers I through VIII by selecting galaxies that meet a range of criteria, including a redshift cut ($0.02 < z < 0.2$) and a half-light radius cut ($1.5 < R_{50}/{\rm kpc} < 10$). We have also, in general, imposed  and $r-$band luminosity ($10^{9.5} < L_r /L_\odot < 10^{11}$), all in the interest of balancing the competing interests of a uniform sample for the stack and sufficiently large numbers. In Paper VIII, we deviated from this luminosity cut because we were investigating the azimuthal dependence of the CGM of massive ellipticals (stellar mass, M$_*$, larger than $10^{10.4}$ M$_\odot$). Now we explore a lower mass primary sample (so we lower the luminosity cut to $10^{9} L_\odot$) and require them to have a stellar mass less than $10^{10.2}$ M$_\odot$, but also greater than $10^{9.0}$ M$_\odot$ to again limit the range of objects in the stack. Decreasing the luminosity criterion to $10^9 $ L$_\odot$ from $10^{9.5}$ L$_\odot$ results in roughly a 10\% increase in the accepted systems. Finally, we also define a minimum ellipticity criterion of $e > 0.25$ \citep{Zhang2022} to ensure that the position angle of the major axis is well-defined. The choice of $e > 0.25$ strikes a balance between larger statistical samples and better measurements, discussed in detail in Paper VIII. After additional selection cuts on the spectra which will be introduced below, we will obtain the final primary galaxy sample.

To perform the sample selection and subsequent analysis, we use measurements of the position angle, the S\'ersic index ($n$), the ellipticity ($e$) and $r$-band absolute magnitude ($M_r$) from the catalog by \citet{simard}.  We adopt measurements of the stellar mass (M$_*$) from  \cite{Kauffmann2003a,Kauffmann2003b} and \cite{Gallazzi}, and star formation rates (SFR) from the MPA-JHU catalog \citep{Tremonti2004, Brinchmann}.  The SFR estimates are aperture corrected to account for the light outside the SDSS/eBOSS fiber aperture (3 arcseconds for SDSS and 2 arcseconds for eBOSS), which only collects $\sim$ 1/3 of the total light for a typical galaxy at the median redshift of the survey \citep[for details see][]{Brinchmann}. The inclusion of the above measurements limits the primary sample to galaxies from the 7th major data release of SDSS (DR7).

\begin{figure}[htbp]
\begin{center}
\includegraphics[width = 0.48 \textwidth]{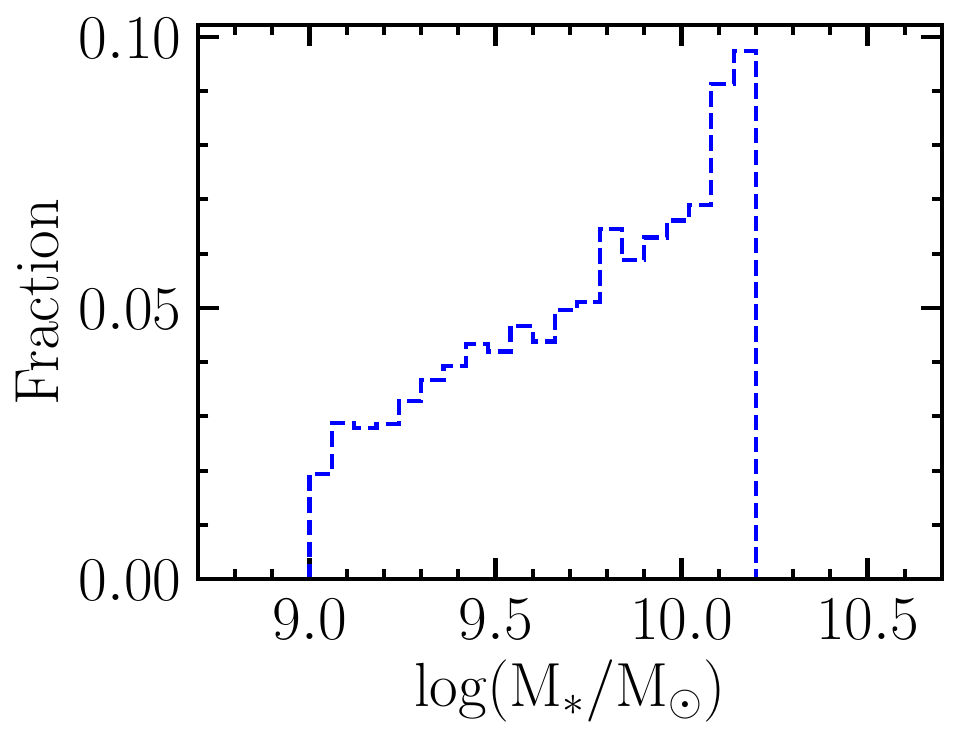}
\end{center}
\caption{The distribution of stellar mass for the primary galaxy sample. We select less massive galaxies by requiring stellar mass to be in the range of  $10^{9.0}$ M$_\odot$ and $10^{10.2}$ M$_\odot$.}  
\label{fig:sm}
\end{figure}

We present the distribution of the galaxy stellar masses in the primary galaxy sample in Figure \ref{fig:sm}.  The mean and median stellar mass of the primary galaxy sample are consistent, $\sim 10^{9.76}$ M$_\odot$ and are more than one order of magnitude lower than those of the massive galaxy sample studied in Paper VIII. As can been seen clearly, the primary galaxy sample shown in Figure \ref{fig:sm} is not complete in this mass range. Fortunately, the Dark Energy Spectroscopic Instrument (DESI) survey \citep{DESI1, DESI2, DESIover, DESI-EDR, DESIvalid} provides potential targets that are more than 2 magnitudes fainter than those provided by SDSS and so, in future work, could help us fill in the lower masses in our primary galaxy sample once consistently derived stellar mass and SFR estimates are available. 

We utilize spectra from both SDSS/eBOSS DR16 \citep{SDSSDR16} and DESI EDR \citep{DESI-EDR} for sightlines projected within a range of scaled projected radii
($0.05 \le r_s \le 0.25)$ for each of our primary galaxies, where $r_s$ is called the scale radius with definition,  $r_s\equiv R_p/R_{\rm vir}$. The DESI EDR data contributes approximately 40\% of the total sightlines (for H$\alpha$ DESI contribnutes 1452 out of the 3631 total). The use of $r_s$ helps to account for the greater than one order of magnitude range in primary galaxy stellar mass. We estimate the primary galaxy's virial radius using the scaling between stellar mass and virial radius described by a high-order polynomial fit to results drawn from the UniverseMachine \citep{Behroozi2019}.
The mean and median virial radius of the  primary galaxy sample is consistent, with a value of 124 kpc. 
In addition to the inner cut in $r_s$,
we have typically also set a lower limit (10 kpc) on the physical projected radius of our sightlines to mitigate possible contamination of the spectra by the central galaxy. Here we decrease that radius to 5 kpc because of our focus on less massive galaxies. We confirm that the results using the inner radius cut of 7.5 kpc are consistent. We attribute this consistency to the 3$\sigma$ flux clipping  we impose, which is described below. 

We briefly describe the calculation of the orientation angle, $\phi$, of each sightline with respect to the major axis of each target galaxy. It is the same as done in Paper VIII. We first calculate the angle on the sky made by the line connecting the center of the primary galaxy and the position of the sightline,
PA$_{1,2}$, using
\begin{equation}
\label{eq:pa}
\tan{(\rm PA_{1,2})} = \frac{\sin(\alpha_1 - \alpha_2)}{\cos \delta_2 \cdot  \tan \delta_1 - \sin \delta_2 \cdot \cos(\alpha_1 - \alpha_2)}
\end{equation}
where $(\alpha_1, \delta_1)$ and $(\alpha_2, \delta_2)$ are the right ascension and declination of the primary galaxy and the sightline. 
We then calculate the difference between PA$_{1,2}$ and the major axis position angle of the target galaxy, restricting the difference to the range of 0$^\circ$ to 90$^\circ$, where 0$^{\circ}$ corresponds to the sightline lying along the major axis (planar) and $90^\circ$ along the minor axis (polar), and refer to the angle as $\phi$ (see two examples in Figure 2 of Paper VIII). 

Our procedure for processing the sightline spectra follows from our previous papers.  For each spectrum, we fit and subtract a 10th-order polynomial to a 300 \AA\ wide section surrounding the observed wavelength of H$\alpha$ at the primary galaxy redshift to remove the continuum.  We then measure the residual H$\alpha$ flux within a velocity window centered on the recessional velocity of the primary galaxy.  
We adopt two velocity windows, $\pm 150$ km s$^{-1}$ and $\pm 210$ km s$^{-1}$, for galaxies with different stellar masses, $10^{9.0} < {\rm M_*/M_\odot} \le 10^{9.7}$ and $10^{9.7} < {\rm M_*/M_\odot} \le 10^{10.2}$ respectively, to broadly track the increasing virial velocities across the mass range.  Requirements on the spectra that are used in the stack include the followings: 1) the continuum level is less than 3 $\times$ $10^{-17}$ erg cm$^{-2}$ s$^{-1}$ \AA$^{-1}$ to limit the noise introduced by the actual spectral target, and 2) the measured emission line flux is within 3$\sigma$ of the mean of the whole sample to remove spectra of interloping strong emitters such as satellite galaxies.  We apply the same procedures and criteria when measuring the H$\beta$, [O {\small III}]$\lambda 5007$ and [N {\small II}]$\lambda 6583$ emission lines except that the maximum allowed continuum  for [O {\small III}]$\lambda 5007$ and H$\beta$ is 2.0 $\times$ $10^{-17}$ erg cm$^{-2}$ s$^{-1}$ \AA$^{-1}$. We present the results of mean flux, and we have confirmed that using the median value produces consistent results (within $2\sigma$) for all bins. There are still two bins whose differences between the mean and median values are approximately $2\sigma$. For the inner, planar bin, the H$\alpha$ and H$\beta$ emission line fluxes have $\lesssim 2\sigma$ differences: the means are $0.0161\pm0.0041$ and $0.0176\pm0.0058$, while the medians are $0.0076\pm0.0040$ and $0.0064\pm0.057$, respectively, in units of $10^{-17}$\,erg\,cm$^{-2}$\,s$^{-1}$\,\AA$^{-1}$. For the outer, polar bin, the [O {\small III}] emission line fluxes are also different by $ \lesssim 2\sigma$: the mean is $0.0028\pm0.0025$ and the median is $0.0084\pm0.0027$ in units of $10^{-17}$\,erg\,cm$^{-2}$\,s$^{-1}$\,\AA$^{-1}$. However, these differences do not qualitatively affect the conclusions we present from the BPT diagram in the following section.

We adopt the bootstrap uncertainty estimation in the mean flux values using random sampling with replacement. We repeat the bootstrap resampling 1,000 times and calculate the mean emission line flux of the resampled data for each iteration. From the distribution of measurements, we quote use the 16.5 and 83.5 percentiles as the uncertainty range.

\section{Results}
\label{sec:results}

In Paper VIII, we investigated the anisotropic distribution of the CGM for massive early-type galaxies with stellar mass greater than $10^{10.4} {\rm M_\odot}$,  to examine the suggestion by \cite{Ignacio2021} that there are physical differences in the CGM of massive early-type galaxies along the polar (along the minor axis) and planar (along the major axis) directions. Indeed, we found that diagnostic line ratios show stronger AGN ionization signatures along the polar direction at small radii than at other angles or radii compared. 

\subsection{Fluxes}

To search for azimuthal variations in our current sample of less massive galaxies, 
we separate our stacked measurements into two bins based on $\phi$, 
$0^\circ \le \phi  < 45^\circ$  (planar direction) and
$45^\circ \le \phi < 90^\circ$ (polar direction), which constrain the CGM properties along the major and minor axes of the target galaxy, respectively. Then we further divide the data into two radial bins, $0.05 < r_s < 0.125$ and $0.125 < r_s < 0.25$.

\begin{deluxetable}{cccrr}
\tablewidth{0pt}
\tablecaption{The mean stacked [O III], H$\alpha$ and [N II] emission flux vs azimuthal angle and radius}
\tablehead{  \colhead{Line} & \colhead{$\langle \phi \rangle$} & \colhead{$\langle r_s \rangle$}\tablenotemark{a} & \colhead{N} & \colhead{$\langle f \rangle$}\\
 &[$^\circ$]&&&\colhead{
 [$10^{-17}$\,erg\,cm$^{-2}$\,s$^{-1}$\,\AA$^{-1}$]}
 }
\startdata
\multirow{4}{*}{H$\beta$} & \multirow{2}{*}{21} & 0.09 & 368 & $0.0176 \pm 0.0058$ \\
& & 0.19 & 1551 & $-0.0006 \pm 0.0024$ \\
& \multirow{2}{*}{68} & 0.09 & 415 & $0.0058 \pm 0.0049$ \\
& & 0.19 & 1452 & $0.0009 \pm 0.0023$ \\ \\
\multirow{4}{*}{[O {\tiny III}]} & \multirow{2}{*}{21} & 0.09 & 378 & $0.0140 \pm 0.0054$ \\
& & 0.19 & 1581 & $0.0029 \pm 0.0025$ \\
& \multirow{2}{*}{68} & 0.09 & 445 & $0.0262 \pm 0.0050$ \\
& & 0.19 & 1456 & $0.0028 \pm 0.0025$ \\ \\
\multirow{4}{*}{H$\alpha$} & \multirow{2}{*}{21} & 0.09 & 368 & $0.0161 \pm 0.0041$ \\
& & 0.19 & 1444 & $0.0052 \pm 0.0018$ \\
& \multirow{2}{*}{68} & 0.09 & 418 & $0.0168 \pm 0.0035$ \\
& & 0.19 & 1401 & $0.0098 \pm 0.0022$ \\ \\
\multirow{4}{*}{[N {\tiny II}]} & \multirow{2}{*}{21} & 0.09 & 374 & $0.0020 \pm 0.0035$ \\
& & 0.19 & 1456 & $-0.0005 \pm 0.0018$ \\
& \multirow{2}{*}{68} & 0.09 & 449 & $0.0114 \pm 0.0034$ \\
& & 0.19 & 1373 & $0.0037 \pm 0.0018$ \\
 \enddata
\label{tab:flux_rs}
\tablenotetext{a}{$r_s$ is  the  ratio  between  the  physical projected separation and the virial radius of the primary galaxy, $r_s \equiv r_p / r_{\rm vir}$.}
\end{deluxetable}

\begin{figure}[htbp]
\begin{center}
\includegraphics[width = 0.48 \textwidth]{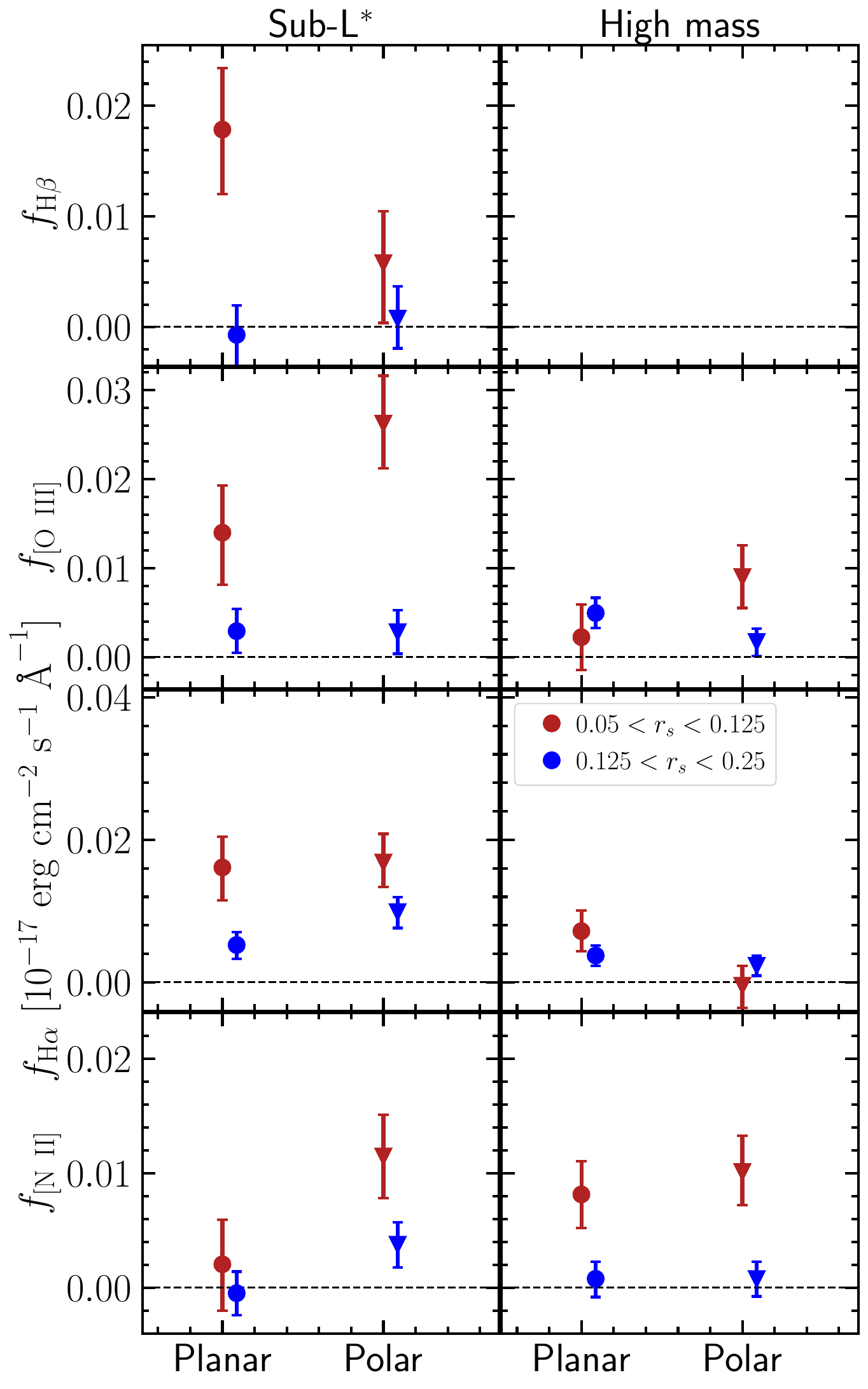}
\end{center}
\caption{The [O {\tiny III}] (top), H$\alpha$ (middle) and [N {\tiny II}] (bottom) emission line flux as a function of azimuthal angle. 
The left column is for the sub-$L^*$ galaxy in this work and the right column is the results of the massive galaxy sample in Paper VIII for easy comparison. The dots represent the emission line fluxes along the major axis (planar) and the triangles for the emission line fluxes along the minor axis (polar). The red symbols represent the emission flux at the inner radii of $0.05 < r_s < 0.125$ and the blue symbols is for the outer radii of $0.125 < r_s < 0.25$. The horizontal dashed line indicates the zero flux for better visualization. There are slight differences in the emission line fluxes among those plotted here for the high mass galaxies and those presented in Paper VIII because we now estimate the virial radius using stellar masses and this affects the radial binning.}  \label{fig:flux_rs}
\end{figure}

We present the measurements of the H$\beta$, [O {\small III}]$\lambda$5007, H$\alpha$ and [N {\small II}]$\lambda$6583 average emission fluxes in each of the two $r_s$ bins as a function of the orientation angle (Table \ref{tab:flux_rs}  and Figure \ref{fig:flux_rs}). Our findings include the following: 
1) a $\sim 3 \sigma$ detection of the H$\beta$ emission line flux in the inner, planar direction and a flux decrement along the azimuthal direction; 2) a significant ($\sim 5\sigma$) detection of [O {\small III}] in the inner polar bin and a flux decrement along the azimuthal direction; 3) a significant detection of H$\alpha$ in all azimuthal and  radial bins (all $> 3 \sigma$) and no evidence for azimuthal variations at either radius; and 4) a significant detection ($> 3.0 \sigma$) of [N {\small II}]  along the polar direction in the inner radial bin.

We notice significant differences in the emission line fluxes between the massive early-type galaxies and the less massive galaxies, including: 1)  a $\sim 3 \sigma$ difference in the H$\alpha$ flux in the polar direction for the inner radii bin;  2) a significant detection ($> 3 \sigma$) of H$\beta$ emission in the less massive galaxies versus non-detection for the massive early-type galaxies for the inner and planar bin; 3) a $\sim 2 \sigma$ difference in the [O {\small III}] flux for the inner and polar bin between the sub-$L^*$ and massive galaxy samples; 4) a comparable [N {\small II}] flux in the polar, inner radial bin for both sub-$L^*$ galaxies and massive early type galaxies. Note that there is no detection of [N {\small II}] emission flux for less massive galaxies in Paper III. The [N {\small II}] emission flux for the inner radius bin is $0.0067\pm0.0048$ in units of $10^{-17}$\,erg\,cm$^{-2}$\,s$^{-1}$\,\AA$^{-1}$ using the spectra of SDSS only, which is consistent with the measurement ($0.0042\pm0.0036$) for the inner radius bin in Paper III. The flux difference is due to the slight different mass range and the usage of scaled radius. And we also find that the emission line fluxes  are mostly greater than those measured for the massive galaxies. This is as expected because we had found a corresponding flux trend with stellar mass in Paper IV, which we attributed to a larger cool gas fraction in lower mass galaxies.

\subsection{Line Ratios}

\begin{figure}[htbp]
\begin{center}
\includegraphics[width = 0.48 \textwidth]{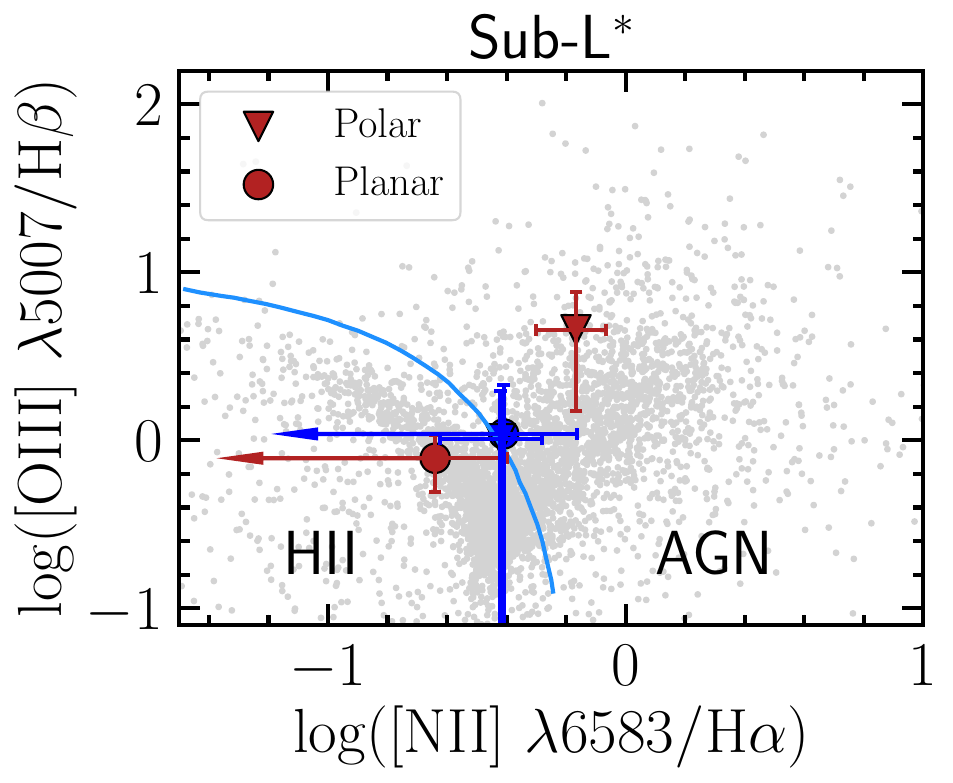}
\includegraphics[width = 0.48 \textwidth]{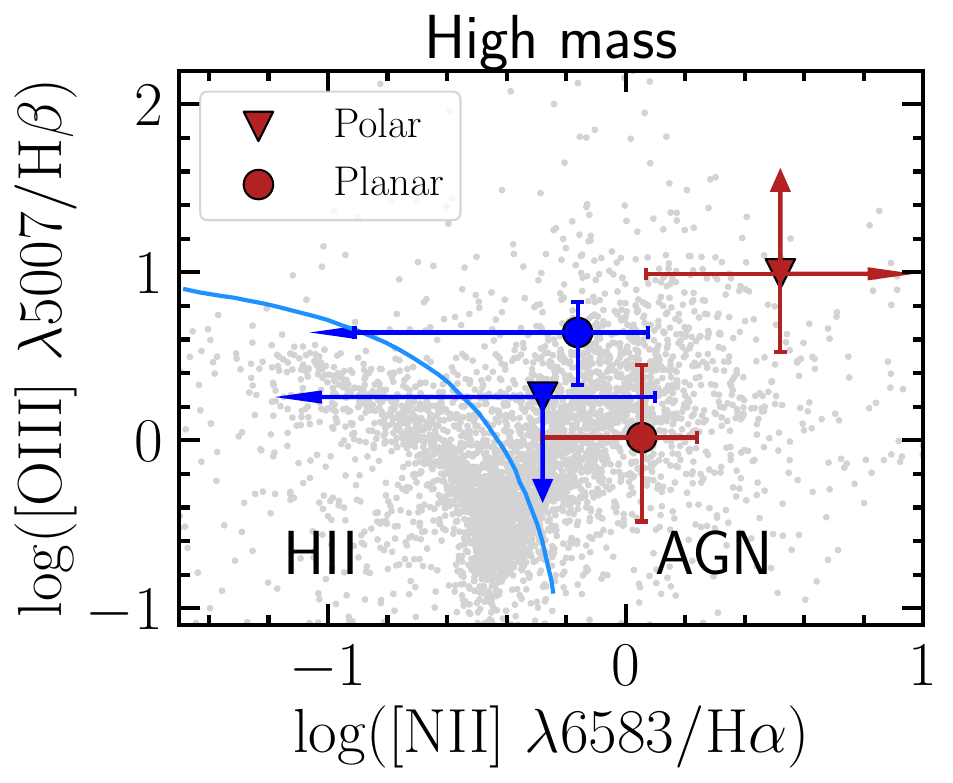}
\end{center}
\caption{The BPT emission line ratios for two azimuthal and two radial bins. These are presented for our lower mass galaxies in the upper panel and reproduced from Paper VIII for the higher mass galaxies in the lower panel. The inner bins, $0.05 < r_s < 0.125$,  are represented in red, while the outer, $0.125 < r_s < 0.25$, are in blue.  The symbols designate the azimuthal bins as given in the legend. The blue curve shows the boundary between the H{\small II} and AGN regions of the diagram and the light grey points represent line ratios of individual SDSS galaxies.}
\label{fig:bpt}
\end{figure}

In addition to the behavior of the fluxes themselves, line ratios are diagnostic. A standard line diagnostic that  compares [N{\small II}]/H$\alpha$ and [O{\small III}]/H$\beta$ and can discriminate between `soft' ionizing sources such as massive stars and `hard' sources such as AGN and shocks is referred to as the BPT diagram \citep{bpt}. 
We present the BPT line ratios, or limits, for the two azimuthal and the two radial bins in Figure \ref{fig:bpt}.
 
One challenge we faced in calculating these ratios is that in some cases we have average flux measurements that are statistically consistent with zero  within their $1 \sigma$ uncertainties). In these cases, we use the $1\sigma$ upper limit when calculating the related line ratio and quote the result as the corresponding limit on the line ratio. 

A second challenge, which we faced in Papers III and VIII, is that the spectral stacks had insufficient S/N to detect H$\beta$. To proceed we adopted H$\beta$/H$\alpha = 0.3$ when calculating the BPT line ratio, a rough value consistent with theoretical expectations \citep{baker, hummer, osterbrock2006}. In the spectral stack of the lower-mass galaxies that are the focus here, we do obtain a statistically significant measurement of H$\beta$ in one of our defined bins. Somewhat concerning given our previous approach and theoretical expectations, we find that the H$\beta$/H$\alpha$ ratio in that bin (the inner, planar bin) is close to one ($1.09 \pm 0.46$), although due to the large uncertainties it is less than 2$\sigma$ discrepant with our assumption that the ratio is 0.3. Although we opt to use our measurements of H$\beta$ when available and limits otherwise, we do confirm that our BPT results are qualitatively unchanged when we revert to our previous approach of adopting a fixed H$\beta$/H$\alpha$ ratio of 0.3.

The line ratios presented in Figure \ref{fig:bpt} lead us to two conclusions.  First, the line ratios along the major axis, at all radii, are consistent with ionization of the CGM by star-formation, as expected for low-mass galaxies ($M_* < 10^{10.4}$ M$_\odot$; Paper III). This is in contrast to the result for the massive elliptical galaxies, where all of the line ratios favored a harder ionizing spectrum. Second, the line ratios along the minor axis favor ionization of the CGM by AGN or shocks, particularly in the inner radial bin. With 99.9\% confidence (see below) we can exclude that the line ratio belongs in the star formation region of the diagram. This result closely follows what we found for the massive elliptical sample in Paper VIII.

To quantitatively assess the statistical significance of the result and determine if the ratio seen in the the inner, polar subsample could plausibly be consistent with those seen in the other three bins, we reconstruct binned averages, randomly drawing with replacement, for the three subsamples that are not the inner polar subsample. The fraction of 10,000 reconstituted samples where the resulting line ratio is as far as that observed for the inner, polar bin (or farther into the AGN regime) occurs only only 4.1\% of the time.  However, this approach combines sightlines across radii. If there are radial differences among sightlines, then we should only compare between inner polar and planar sightlines. When drawing solely from the inner, planar bin, with replacement, and repeating the test we find that we can reproduce or exceed the inner polar ratio only 0.01\% of the time. As such, we conclude that the inner, polar bin differs from the inner, planar bin with confidence of 99.99\%. 

One concern here is that we could be misled because the tests just described are {\sl a  posteori} statistical tests. To address this concern, we construct a control sample and apply the same test.
As in Paper IV, we `move' each primary galaxies to a blank sky position, assign a random intrinsic position angle to the primary galaxy. We repeat our analysis, defining each of the four bins as the `different' bin and drawing comparison samples from the other three bins. In no combination is the sole bin found to be different than the other bins with greater than 1$\sigma$ confidence in the BPT diagram.

\section{Discussion}

There are a number of complicating factors in our measurements of the CGM and the interpretation of possible physical influences that we now discuss.

\subsection{SDSS vs. DESI}

Although the SDSS \citep{SDSS2000} and the DESI surveys \citep{DESI1, DESI2} are comparable in terms of the survey strategy,  sky coverage and spectral resolution,  differences in the reduction pipeline might lead to subtle differences in the spectra that might affect the stacked spectra. Here we examine results from the inner radius bins separately from SDSS and DESI. Our question is whether the result that the inner, polar bin is anomalous relative to the others is caused by an unknown systematic difference between the SDSS and DESI spectra. 

\begin{figure}[htbp]
\begin{center}
\includegraphics[width = 0.48 \textwidth]{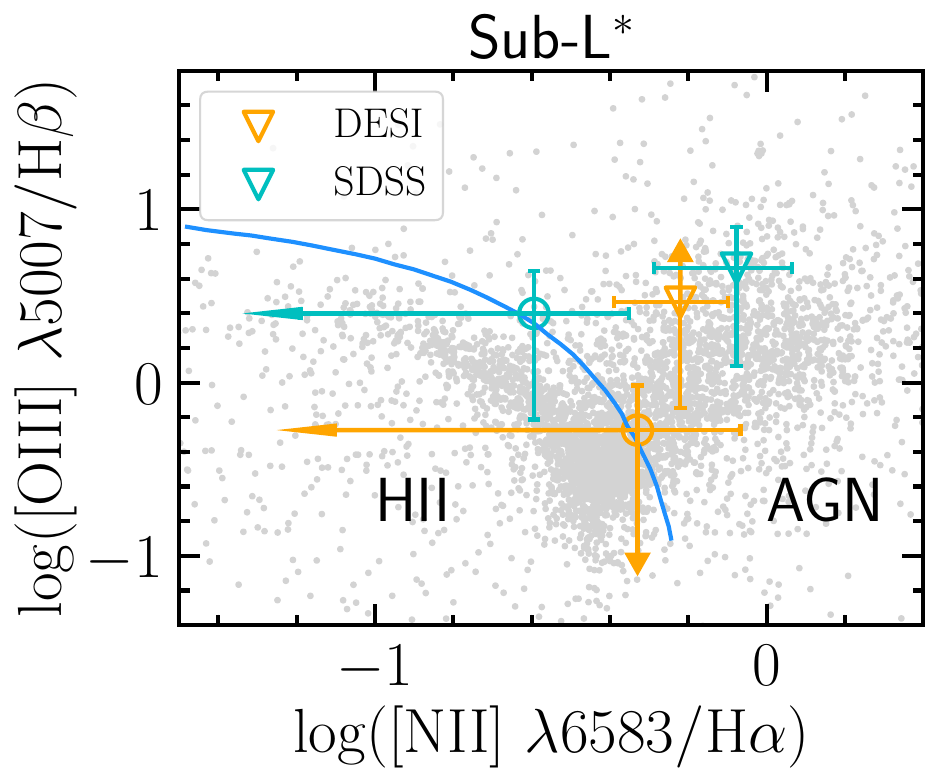}
\end{center}
\caption{The BPT emission line ratios for the two azimuthal bins at the inner radii.  The open triangles and circles represent the polar and planar bins, respectively. Results in  orange and cyan use stacked DESI and SDSS spectra, respectively. As in Figure \ref{fig:bpt}, the blue curve shows the boundary between the H II and AGN regions of the diagram and the light grey points represent line ratios of individual SDSS galaxies.}
\label{fig:sepBPT}
\end{figure}

In Figure \ref{fig:sepBPT} we present the line ratios for the inner radius bins in which the measurements using the spectra of  SDSS and DESI separately.  Although the error bars are larger because of smaller data sample, and hence the distinction between the polar and planar results are less significant, the sense of the result (polar lying in the AGN region, planar in the star forming region) is reproduced in each subset. We conclude that our finding is not due to a distinct systematic error in either the SDSS or DESI samples.

\subsection{Dependence on galaxy properties}

In Papers II/III/IV, we found that the emission line flux from the CGM correlates with the stellar mass, SFR, and morphology of the galaxy. As such, it is always a danger when comparing samples selected for one reason that any observed differences might instead originate from these differences in galaxy properties. Although we cannot envision reasons why selecting on azimuth or radius of the sightlines would connect to the properties of the primary galaxy, we nevertheless test that the primaries in the various subsamples are all similar. We find that the samples are nearly indistinguishable, with mean stellar masses for the four subsamples ranging between 
$10^{9.78}$ M$_\odot$ and $10^{9.83}$ M$_\odot$ with a difference between mean and median values that is less than 0.1 dex, the S\'ersic $n$ indices between 3.12 and 3.45 with a difference between mean and median value that is less than 0.2, and the SFRs between 0.55 and 0.62 M$_\odot$/yr with a difference between mean and median value that is less than 0.1 dex. 
The mean and median stellar masses and S\'ersic $n$ indices are consistent between the subsamples probed by SDSS and DESI, and mean and median values of the SFR of the DESI subsample are $\sim$ 0.1 dex lower than those of the SDSS subsample. Based on previous results, Papers II/III/IV, variations within these ranges are not expected to lead to detectable emission line differences.

A different property that could affect the line ratios is metallicity. For example, the [N {\small II}]$\lambda 6583$/H$\alpha$ ratio is known to correlate strongly with both metallicity and ionization parameter \citep{Levesque2010}. As such, the leftward shift of the data in Figure \ref{fig:bpt} for the lower mass galaxies relative to the higher mass ones is plausibly related to a metallicity difference in the CGM gas in these two galaxy samples. Although we do not have metallicity measurements for our samples and gas metallicity estimates are still rough in other samples, \cite{prochaska2017} estimated the CGM metallicity to be $-0.51$ for $L^*$ galaxies
and \cite{Berg2019} estimated
the CGM metallicity of star-forming galaxies with stellar mass $< 10^{10.5}$ M$_\odot$ to be lower ($-1.75 < [X/H] < -0.5$ ). Because lower metallicity corresponds to lower [N {\small II}]$\lambda 6583$/H$\alpha$ ratio \citep{Levesque2010}, the sense of the effect is consistent.

Regarding the polar-planar asymmetry we find, at least at inner radii, there are two reasons why this is unlikely to be a result of metallicity differences. First
\cite{Pointon2019} found no obvious difference in the CGM metallicity ([Si/H]) along the major and minor axis of typical $L^*$ galaxies. Second, in none of the models presented by \cite{Levesque2010}, which examined line ratios for star forming galaxies, do they find BPT line ratios well into the active galaxy regime. The inferred hardening of the radiation needed to produce the inner, polar measurement cannot arise from a metallicity difference relative to the gas in the other bins.


\subsection{Extragalactic Ultraviolet (UV) Background}

Although the motivation for this study is based on understanding the contribution of different sources of ionizing radiation from within the galaxy, the CGM is also susceptible to ionization by the extragalactic UV field, alternatively referred to as the ultraviolet background \citep[UVB;][]{Haardt1996, Haardt2001, Haardt2012}. The CGM gas traced by QSO absorption lines is typically assumed to be photoionized primarily by the UVB, especially beyond 50 kpc \citep[eg., ][]{Fumagalli2011, Werk2014}. 

Results we have obtained using emission lines suggest that the gas within 50 kpc is ionized principally by sources within the galaxy. In this study alone, the change in BPT line ratios along in the polar, inner bin suggests not only that galaxy orientation plays a role, but that the effect is strongest closer to the galaxy. Likewise, the finding that the nature of the BPT ratios depends on galaxy mass (Paper III) indicates the role that the galaxies themselves play in ionizing the CGM gas, at least within 50 kpc which is the focus of our studies.
 

\subsection{Comparison to Simulations}
\label{sec:simu}

Connecting this result to simulations is not straightforward because line ratios are not synthesized. However, there are many results that suggest polar asymmetries should exist.
For example, \cite{Mitchell2020} studied the properties of outflows on galactic scale using EAGLE simulations \citep{EAGLE1,EAGLE2} and found 
a bimodal outflow distribution that aligns with the galaxy minor axis at $r > 0.1 R_{\rm vir}$ when selecting gas particles with radial velocities $v_{\rm rad} > 0.25 V_{\rm max}$
for a Milky Way-like galaxy. 
\cite{Peroux2020} found that 
the CGM mass flow rate and metallicity correlate with azimuthal angle across a  
wide range of stellar mass (8.5 $< \log M_* / M_\odot <$ 10.5) using the high-resolution TNG50 cosmological magnetohydrodynamical simulation \citep{Nelson2019, Pillepich2019}.
\cite{Truong2021} predicted that the CGM is anisotropic in its thermodynamical properties and chemical content over a large galaxy mass range.  
Finally, \cite{Ramesh2023a} showed that magnetic fields in the CGM of TNG50 galaxies have significant angular structure, and that gas along the minor axes of galaxies is more strongly affected, especially for the galaxy with $M_* \sim 10^{10}$ M$_\odot$.  They also found that the anisotropy arises as a result of two different feedback processes (stellar feedback for low-mass systems and AGN feedback for high-mass systems), qualitatively consistent with the broad trend in line ratios that we find.

It is clearly seen that the signature of a harder ionizing spectrum is only found at small radii along the polar direction for the stellar mass range we are able to explore. This phenomena might indicate that the feedback is trapped at the inner CGM. Simulations could help to further investigate this phenomena.

In any case,  drawing physical inferences from the comparison of  simulations to the results from stacking tens of thousands of spectra is challenging. The CGM in each galaxy is a baryon reservoir with multi-phase, multi-ionization-state and  geometrically-complex structure \citep{CGM2017, CGM2023}. It is not evident how those add up to produce a mean line flux in the stacking analysis. Furthermore, the interplay of supernovae and possible AGN feedback in galaxies makes it even more challenging to disentangle the CGM properties from the stack results.   Instead, we await the detection of emission lines in individual, normal, low-redshift galaxies where differences among individual galaxies may help break some of these degeneracies. We are encouraged that some studies of the CGM of individual nearby galaxies already exist \citep{Herenz2023, Nielsen2023}, the sample is still small and the galaxies studied are not representative. 
For the warm CGM such observations may come relatively sooner using ultraviolet emission lines observed with satellites like {\sl Aspera} \citep{aspera}, while for the hot CGM they may eventually be available from X-ray facilities such as the Hot Universe Baryon Surveyor \citep[HUBS,][]{HUBS2020} or the Line Emission Mapper \citep[LEM,][]{Kraft2022, Khabibullin2023}.

\section{Summary}
\label{sec:sum}

To provide further constraints on models of the feedback mechanism and its effect on the circumgalactic medium, we applied a methodology developed to measure the emission line fluxes from the cool ($T\sim 10^4$K) CGM specifically to lower-mass galaxies. We selected primary galaxies with stellar mass in the range of $10^{9.0}<$ M$_* < 10^{10.2}$ M$_\odot$ and  ellipticity $e > 0.25$. We calculated the relative position angle between each sightline and the major axis of the associated central galaxy, and then combined the H$\beta$, [O {\small III}], H$\alpha$ and [N {\small II}] emission fluxes to investigate trends with scaled projected radii and azimuthal angle.

In contrast to massive early-type galaxies where H$\alpha$ emission line flux significantly drops from the major to the minor axis, the flux remains unchanged at different azimuthal angles for the inner radial bin of sub-$L^*$ galaxies. Notably, the [N{\small II}] emission line flux is detected with a significance of $> 3.0 \sigma$ in the polar direction for the inner radial bin, despite previous challenges in detecting [N {\small II}] emission in sub-$L^*$ galaxies. All fluxes exhibit significant radial declines except for the [N {\small II}] flux in the planar direction. Comparing sub-$L^*$ and massive galaxies, we observe that sub-$L^*$ galaxies have higher emission line fluxes in general and 
describe differences in mean line fluxes for specific emission lines.


To follow on our earlier finding that hardness of the ionization spectrum has an azimuthal asymmetry in massive elliptical galaxies (Paper VIII), we explore the BPT diagnostic line ratio for the CGM along the major and minor axis of the central galaxy for sub-$L^*$ galaxies. We have detection of H$\beta$ emission line flux for the inner radii bins due to a higher cool gas fraction in lower mass galaxies and the increased data sample, and we quote limits for the logarithm of the ratios when either the numerator or denominator is consistent with zero to within 1$\sigma$. With the anticipated increase in the number of spectra from DESI \citep[][]{DESI1, DESI2} by almost two orders of magnitude compared to that of SDSS, we expect to detect H$\beta$ emission flux for more data bins by stacking with reasonable signal to noise ratio.  The ratios for the combined spectra in the inner, polar bin are statistically different from those in any other radial or azimuthal bin (confidence between 95.9\% and 99.9\% depending on the test). 
The sense of the result, that the inner, polar bin exhibits the signature of a harder ionizing spectrum is consistent with the result found for massive elliptical galaxies (Paper VIII). The hardening of the spectrum at small radii along the polar axis appears to be a ubiquitous feature. Future simulations could investigate the direction and distribution of outflows to further understand the details of feedback mechanism.

\section{Acknowledgments}

HZ acknowledges financial support from the start-up funding of the Huazhong University of Science and Technology and the National Science Foundation of China grant (No. 12303007). ML acknowledges the support from the National Key Research and Development Program of China (2022YFA1602903), from the National Science Foundation of China (12147103, 12273010), and from the Fundamental Research Funds for the Central Universities(226-2022-00216). DZ acknowledges financial support from NSF grant AST-2006785. The authors thank the referee for comments that helped us improve the manuscript. This research was supported by the Munich Institute for Astro-, Particle and BioPhysics (MIAPbP) which is funded by the Deutsche Forschungsgemeinschaft (DFG, German Research Foundation) under Germany's Excellence Strategy EXC-2094 390783311. The authors gratefully acknowledge the SDSS III team and DESI team for providing a valuable resource to the community.
Funding for SDSS-III has been provided by the Alfred P. Sloan Foundation, the Participating I institutions, the National Science Foundation, and the U.S. Department of Energy Office of Science. The SDSS-III web site is http://www.sdss3.org/.

SDSS-III is managed by the Astrophysical Research Consortium for the Participating Institutions of the SDSS-III Collaboration including the University of Arizona, the Brazilian Participation Group, Brookhaven National Laboratory, Carnegie Mellon University, University of Florida, the French Participation Group, the German Participation Group, Harvard University, the Instituto de Astrofisica de Canarias, the Michigan State/Notre Dame/JINA Participation Group, Johns Hopkins University, Lawrence Berkeley National Laboratory, Max Planck Institute for Astrophysics, Max Planck Institute for Extraterrestrial Physics, New Mexico State University, New York University, Ohio State University, Pennsylvania State University, University of Portsmouth, Princeton University, the Spanish Participation Group, University of Tokyo, University of Utah, Vanderbilt University, University of Virginia, University of Washington, and Yale University.

This research used data obtained with the Dark Energy Spectroscopic Instrument (DESI). DESI construction and operations is managed by the Lawrence Berkeley National Laboratory. This material is based upon work supported by the U.S. Department of Energy, Office of Science, Office of High-Energy Physics, under Contract No. DE–AC02–05CH11231, and by the National Energy Research Scientific Computing Center, a DOE Office of Science User Facility under the same contract. Additional support for DESI was provided by the U.S. National Science Foundation (NSF), Division of Astronomical Sciences under Contract No. AST-0950945 to the NSF’s National Optical-Infrared Astronomy Research Laboratory; the Science and Technology Facilities Council of the United Kingdom; the Gordon and Betty Moore Foundation; the Heising-Simons Foundation; the French Alternative Energies and Atomic Energy Commission (CEA); the National Council of Science and Technology of Mexico (CONACYT); the Ministry of Science and Innovation of Spain (MICINN), and by the DESI Member Institutions: www.desi.lbl.gov/collaborating-institutions. The DESI collaboration is honored to be permitted to conduct scientific research on Iolkam Du’ag (Kitt Peak), a mountain with particular significance to the Tohono O’odham Nation. Any opinions, findings, and conclusions or recommendations expressed in this material are those of the author(s) and do not necessarily reflect the views of the U.S. National Science Foundation, the U.S. Department of Energy, or any of the listed funding agencies.

\bibliography{bibliography}

\begin{thebibliography}{98}
\expandafter\ifx\csname natexlab\endcsname\relax\def\natexlab#1{#1}\fi

\bibitem[{{Ahumada} {et~al}\mbox{.}(2020){Ahumada}, {Allende Prieto},
  {Almeida}, {Anders}, \& {Anderson}}]{SDSSDR16}
{Ahumada} R., {Allende Prieto} C., {Almeida} A., {Anders} F., {Anderson}, 2020,
  \apjs, 249, 3

\bibitem[{{Alam} {et~al}\mbox{.}(2015){Alam}, {Albareti}, {Allende Prieto},
  {Anders}, {Anderson}, {Anderton}, {Andrews}, \& et~al.}]{SDSS12}
{Alam} S., {Albareti} F.~D., {Allende Prieto} C., {Anders} F., {Anderson}
  S.~F., {Anderton} T., {Andrews} B.~H., et~al., 2015, \apjs, 219, 12

\bibitem[{{Baker} \& {Menzel}(1938)}]{baker}
{Baker} J.~G., {Menzel} D.~H., 1938, \apj, 88, 52

\bibitem[{{Baldwin}, {Phillips} \& {Terlevich}(1981){Baldwin}, {Phillips}, \&
  {Terlevich}}]{bpt}
{Baldwin} J.~A., {Phillips} M.~M., {Terlevich} R., 1981, \pasp, 93, 5

\bibitem[{{Behroozi} {et~al}\mbox{.}(2019){Behroozi}, {Wechsler}, {Hearin}, \&
  {Conroy}}]{Behroozi2019}
{Behroozi} P., {Wechsler} R.~H., {Hearin} A.~P., {Conroy} C., 2019, \mnras,
  488, 3143

\bibitem[{{Berg} {et~al}\mbox{.}(2019){Berg}, {Howk}, {Lehner}, {Wotta},
  {O'Meara}, {Bowen}, {Burchett}, {Peeples}, \& {Tejos}}]{Berg2019}
{Berg} M.~A. {et~al.}, 2019, \apj, 883, 5

\bibitem[{{Bordoloi} {et~al}\mbox{.}(2011){Bordoloi}, {Lilly}, {Knobel},
  {Bolzonella}, {Kampczyk}, {Carollo}, {Iovino}, {Zucca}, {Contini}, {Kneib},
  {Le Fevre}, {Mainieri}, {Renzini}, {Scodeggio}, {Zamorani}, {Balestra},
  {Bardelli}, {Bongiorno}, {Caputi}, {Cucciati}, {de la Torre}, {de Ravel},
  {Garilli}, {Kova{\v{c}}}, {Lamareille}, {Le Borgne}, {Le Brun}, {Maier},
  {Mignoli}, {Pello}, {Peng}, {Perez Montero}, {Presotto}, {Scarlata},
  {Silverman}, {Tanaka}, {Tasca}, {Tresse}, {Vergani}, {Barnes}, {Cappi},
  {Cimatti}, {Coppa}, {Diener}, {Franzetti}, {Koekemoer}, {L{\'o}pez-Sanjuan},
  {McCracken}, {Moresco}, {Nair}, {Oesch}, {Pozzetti}, \&
  {Welikala}}]{Bordoloi2011}
{Bordoloi} R. {et~al.}, 2011, \apj, 743, 10

\bibitem[{{Brinchmann} {et~al}\mbox{.}(2004){Brinchmann}, {Charlot}, {White},
  {Tremonti}, {Kauffmann}, {Heckman}, \& {Brinkmann}}]{Brinchmann}
{Brinchmann} J., {Charlot} S., {White} S.~D.~M., {Tremonti} C., {Kauffmann} G.,
  {Heckman} T., {Brinkmann} J., 2004, MNRAS, 351, 1151

\bibitem[{{Cai} {et~al}\mbox{.}(2017){Cai}, {Fan}, {Yang}, {Bian}, {Prochaska},
  {Zabludoff}, {McGreer}, {Zheng}, {Green}, {Cantalupo}, {Frye}, {Hamden},
  {Jiang}, {Kashikawa}, \& {Wang}}]{Cai2017}
{Cai} Z. {et~al.}, 2017, \apj, 837, 71

\bibitem[{{Ceverino} \& {Klypin}(2009)}]{Ceverino2009}
{Ceverino} D., {Klypin} A., 2009, \apj, 695, 292

\bibitem[{{Chen}(2017)}]{Chen2017a}
{Chen} H.-W., 2017, Astrophysics and Space Science Library, Vol. 430, {The
  Circumgalactic Medium in Massive Halos}, {Fox} A., {Dav{\'e}} R., eds., p.
  167

\bibitem[{{Chen} {et~al}\mbox{.}(2019){Chen}, {Boettcher}, {Johnson}, {Zahedy},
  {Rudie}, {Cooksey}, {Rauch}, \& {Mulchaey}}]{Chen2019}
{Chen} H.-W., {Boettcher} E., {Johnson} S.~D., {Zahedy} F.~S., {Rudie} G.~C.,
  {Cooksey} K.~L., {Rauch} M., {Mulchaey} J.~S., 2019, \apjl, 878, L33

\bibitem[{{Chung} {et~al}\mbox{.}(2021){Chung}, {Vargas}, {Hamden}, {McMahon},
  {Gonzales}, {Khan}, {Agarwal}, {Bailey}, {Behroozi}, {Brendel}, {Choi},
  {Connors}, {Corlies}, {Corliss}, {Dettmar}, {Dolana}, {Douglas}, {Guzman},
  {Hamara}, {Harris}, {Harshman}, {Hergenrother}, {Hoadley}, {Kidd}, {Kim},
  {Li}, {Montoya}, {Sauve}, {Schiminovich}, {Selznick}, {Siegmund}, {Ward},
  {Wolcott}, \& {Zaritsky}}]{aspera}
{Chung} H. {et~al.}, 2021, in Society of Photo-Optical Instrumentation
  Engineers (SPIE) Conference Series, Vol. 11819, Society of Photo-Optical
  Instrumentation Engineers (SPIE) Conference Series, p. 1181903

\bibitem[{{Crain} {et~al}\mbox{.}(2015){Crain}, {Schaye}, {Bower}, {Furlong},
  {Schaller}, {Theuns}, {Dalla Vecchia}, {Frenk}, {McCarthy}, {Helly},
  {Jenkins}, {Rosas-Guevara}, {White}, \& {Trayford}}]{EAGLE1}
{Crain} R.~A. {et~al.}, 2015, \mnras, 450, 1937

\bibitem[{{Croft} {et~al}\mbox{.}(2016){Croft}, {Miralda-Escud{\'e}}, {Zheng},
  {Bolton}, {Dawson}, {Peterson}, {York}, \& et~al}]{croft2016}
{Croft} R.~A.~C., {Miralda-Escud{\'e}} J., {Zheng} Z., {Bolton} A., {Dawson}
  K.~S., {Peterson} J.~B., {York} D.~G., et~al, 2016, \mnras, 457, 3541

\bibitem[{{Cui} {et~al}\mbox{.}(2020){Cui}, {Chen}, {Gao}, {Guo}, {Jin},
  {Wang}, {Wang}, {Wang}, {Wang}, {Wang}, {Wang}, {Yuan}, \&
  {Zhang}}]{HUBS2020}
{Cui} W. {et~al.}, 2020, Journal of Low Temperature Physics, 199, 502

\bibitem[{{DESI Collaboration} {et~al}\mbox{.}(2022){DESI Collaboration},
  {Abareshi}, {Aguilar}, {Ahlen}, {Alam}, {Alexander}, {Alfarsy}, {Allen},
  {Allende Prieto}, {Alves}, \& et~al.}]{DESIover}
{DESI Collaboration} {et~al.}, 2022, \aj, 164, 207

\bibitem[{{DESI Collaboration} {et~al}\mbox{.}(2023){DESI Collaboration},
  {Adame}, {Aguilar}, {Ahlen}, {Alam}, {Aldering}, {Alexander}, {Alfarsy},
  {Allende Prieto}, {Alvarez}, {Alves}, {Anand}, {Andrade-Oliveira},
  {Armengaud}, {Asorey}, {Avila}, {Aviles}, {Bailey},
  {Balaguera-Antol{\'\i}nez}, {Ballester}, {Baltay}, {Bault}, {Bautista},
  {Behera}, {Beltran}, {BenZvi}, {Beraldo e Silva}, {Bermejo-Climent}, {Berti},
  {Besuner}, {Beutler}, {Bianchi}, {Blake}, {Blum}, {Bolton}, {Brieden},
  {Brodzeller}, {Brooks}, {Brown}, {Buckley-Geer}, {Burtin}, {Cabayol-Garcia},
  {Cai}, {Canning}, {Cardiel-Sas}, {Carnero Rosell}, {Castander},
  {Cervantes-Cota}, {Chabanier}, {Chaussidon}, {Chaves-Montero}, {Chen},
  {Chuang}, {Claybaugh}, {Cole}, {Cooper}, {Cuceu}, {Davis}, {Dawson}, {de
  Belsunce}, {de la Cruz}, {de la Macorra}, {de Mattia}, {Demina},
  {Demirbozan}, {DeRose}, {Dey}, {Dey}, {Dhungana}, {Ding}, {Ding}, {Doel},
  {Doshi}, {Douglass}, {Edge}, {Eftekharzadeh}, {Eisenstein}, {Elliott},
  {Escoffier}, {Fagrelius}, {Fan}, {Fanning}, {Fawcett}, {Ferraro}, {Ereza},
  {Flaugher}, {Font-Ribera}, {Forero-S{\'a}nchez}, {Forero-Romero}, {Frenk},
  {G{\"a}nsicke}, {Garc{\'\i}a}, {Garc{\'\i}a-Bellido}, {Garcia-Quintero},
  {Garrison}, {Gil-Mar{\'\i}n}, {Golden-Marx}, {Gontcho}, {Gonzalez-Morales},
  {Gonzalez-Perez}, {Gordon}, {Graur}, {Green}, {Gruen}, {Guy}, {Hadzhiyska},
  {Hahn}, {Han}, {Hanif}, {Herrera-Alcantar}, {Honscheid}, {Hou}, {Howlett},
  {Huterer}, {Ir{\v{s}}i{\v{c}}}, {Ishak}, {Jacques}, {Jana}, {Jiang},
  {Jimenez}, {Jing}, {Joudaki}, {Jullo}, {Juneau}, {Kizhuprakkat},
  {Kara{\c{c}}ayl{\i}}, {Karim}, {Kehoe}, {Kent}, {Khederlarian}, {Kim},
  {Kirkby}, {Kisner}, {Kitaura}, {Kneib}, {Koposov}, {Kov{\'a}cs}, {Kremin},
  {Krolewski}, {L'Huillier}, {Lambert}, {Lamman}, {Lan}, {Landriau}, {Lang},
  {Lange}, {Lasker}, {Le Guillou}, {Leauthaud}, {Levi}, {Li}, {Linder},
  {Lyons}, {Magneville}, {Manera}, {Manser}, {Margala}, {Martini}, {McDonald},
  {Medina}, {Medina-Varela}, {Meisner}, {Mena-Fern{\'a}ndez}, {Meneses-Rizo},
  {Mezcua}, {Miquel}, {Montero-Camacho}, {Moon}, {Moore}, {Moustakas},
  {Mueller}, {Mundet}, {Mu{\~n}oz-Guti{\'e}rrez}, {Myers}, {Nadathur},
  {Napolitano}, {Neveux}, {Newman}, {Nie}, {Nikutta}, {Niz}, {Norberg},
  {Noriega}, {Paillas}, {Palanque-Delabrouille}, {Palmese}, {Zhiwei},
  {Parkinson}, {Penmetsa}, {Percival}, {P{\'e}rez-Fern{\'a}ndez},
  {P{\'e}rez-R{\`a}fols}, {Pieri}, {Poppett}, {Porredon}, {Pothier}, {Prada},
  {Pucha}, {Raichoor}, {Ram{\'\i}rez-P{\'e}rez}, {Ramirez-Solano},
  {Rashkovetskyi}, {Ravoux}, {Rocher}, {Rockosi}, {Ross}, {Rossi}, {Ruggeri},
  {Ruhlmann-Kleider}, {Sabiu}, {Said}, {Saintonge}, {Samushia}, {Sanchez},
  {Saulder}, {Schaan}, {Schlafly}, {Schlegel}, {Scholte}, {Schubnell}, {Seo},
  {Shafieloo}, {Sharples}, {Sheu}, {Silber}, {Sinigaglia}, {Siudek}, {Slepian},
  {Smith}, {Sprayberry}, {Stephey}, {Su{\'a}rez-P{\'e}rez}, {Sun}, {Tan},
  {Tarl{\'e}}, {Tojeiro}, {Ure{\~n}a-L{\'o}pez}, {Vaisakh}, {Valcin}, {Valdes},
  {Valluri}, {Vargas-Maga{\~n}a}, {Variu}, {Verde}, {Walther}, {Wang}, {Wang},
  {Weaver}, {Weaverdyck}, {Wechsler}, {White}, {Xie}, {Yang}, {Y{\`e}che},
  {Yu}, {Yuan}, {Zhang}, {Zhang}, {Zhao}, {Zheng}, {Zhou}, {Zhou}, {Zou},
  {Zou}, \& {Zu}}]{DESI-EDR}
{DESI Collaboration} {et~al.}, 2023, arXiv e-prints, arXiv:2306.06308

\bibitem[{{DESI Collaboration} {et~al}\mbox{.}(2024){DESI Collaboration},
  {Adame}, {Aguilar}, {Ahlen}, {Alam}, {Aldering}, {Alexander}, {Alfarsy},
  {Allende Prieto}, {Alvarez}, \& et~al.}]{DESIvalid}
{DESI Collaboration} {et~al.}, 2024, \aj, 167, 62

\bibitem[{{DESI Collaboration} {et~al}\mbox{.}(2016{\natexlab{a}}){DESI
  Collaboration}, {Aghamousa}, {Aguilar}, {Ahlen}, {Alam}, {Allen}, {Allende
  Prieto}, {Annis}, {Bailey}, {Balland}, {Ballester}, {Baltay}, {Beaufore},
  {Bebek}, {Beers}, {Bell}, {Bernal}, {Besuner}, {Beutler}, {Blake}, {Bleuler},
  {Blomqvist}, {Blum}, {Bolton}, {Briceno}, {Brooks}, {Brownstein},
  {Buckley-Geer}, {Burden}, {Burtin}, {Busca}, {Cahn}, {Cai}, {Cardiel-Sas},
  {Carlberg}, {Carton}, {Casas}, {Castander}, {Cervantes-Cota}, {Claybaugh},
  {Close}, {Coker}, {Cole}, {Comparat}, {Cooper}, {Cousinou}, {Crocce}, {Cuby},
  {Cunningham}, {Davis}, {Dawson}, {de la Macorra}, {De Vicente}, {Delubac},
  {Derwent}, {Dey}, {Dhungana}, {Ding}, {Doel}, {Duan}, {Ealet}, {Edelstein},
  {Eftekharzadeh}, {Eisenstein}, {Elliott}, {Escoffier}, {Evatt}, {Fagrelius},
  {Fan}, {Fanning}, {Farahi}, {Farihi}, {Favole}, {Feng}, {Fernandez},
  {Findlay}, {Finkbeiner}, {Fitzpatrick}, {Flaugher}, {Flender}, {Font-Ribera},
  {Forero-Romero}, {Fosalba}, {Frenk}, {Fumagalli}, {Gaensicke}, {Gallo},
  {Garcia-Bellido}, {Gaztanaga}, {Pietro Gentile Fusillo}, {Gerard},
  {Gershkovich}, {Giannantonio}, {Gillet}, {Gonzalez-de-Rivera},
  {Gonzalez-Perez}, {Gott}, {Graur}, {Gutierrez}, {Guy}, {Habib}, {Heetderks},
  {Heetderks}, {Heitmann}, {Hellwing}, {Herrera}, {Ho}, {Holland}, {Honscheid},
  {Huff}, {Hutchinson}, {Huterer}, {Hwang}, {Illa Laguna}, {Ishikawa},
  {Jacobs}, {Jeffrey}, {Jelinsky}, {Jennings}, {Jiang}, {Jimenez}, {Johnson},
  {Joyce}, {Jullo}, {Juneau}, {Kama}, {Karcher}, {Karkar}, {Kehoe}, {Kennamer},
  {Kent}, {Kilbinger}, {Kim}, {Kirkby}, {Kisner}, {Kitanidis}, {Kneib},
  {Koposov}, {Kovacs}, {Koyama}, {Kremin}, {Kron}, {Kronig}, {Kueter-Young},
  {Lacey}, {Lafever}, {Lahav}, {Lambert}, {Lampton}, {Landriau}, {Lang},
  {Lauer}, {Le Goff}, {Le Guillou}, {Le Van Suu}, {Lee}, {Lee}, {Leitner},
  {Lesser}, {Levi}, {L'Huillier}, {Li}, {Liang}, {Lin}, {Linder}, {Loebman},
  {Luki{\'c}}, {Ma}, {MacCrann}, {Magneville}, {Makarem}, {Manera}, {Manser},
  {Marshall}, {Martini}, {Massey}, {Matheson}, {McCauley}, {McDonald},
  {McGreer}, {Meisner}, {Metcalfe}, {Miller}, {Miquel}, {Moustakas}, {Myers},
  {Naik}, {Newman}, {Nichol}, {Nicola}, {Nicolati da Costa}, {Nie}, {Niz},
  {Norberg}, {Nord}, {Norman}, {Nugent}, {O'Brien}, {Oh}, {Olsen}, {Padilla},
  {Padmanabhan}, {Padmanabhan}, {Palanque-Delabrouille}, {Palmese},
  {Pappalardo}, {P{\^a}ris}, {Park}, {Patej}, {Peacock}, {Peiris}, {Peng},
  {Percival}, {Perruchot}, {Pieri}, {Pogge}, {Pollack}, {Poppett}, {Prada},
  {Prakash}, {Probst}, {Rabinowitz}, {Raichoor}, {Ree}, {Refregier}, {Regal},
  {Reid}, {Reil}, {Rezaie}, {Rockosi}, {Roe}, {Ronayette}, {Roodman}, {Ross},
  {Ross}, {Rossi}, {Rozo}, {Ruhlmann-Kleider}, {Rykoff}, {Sabiu}, {Samushia},
  {Sanchez}, {Sanchez}, {Schlegel}, {Schneider}, {Schubnell}, {Secroun},
  {Seljak}, {Seo}, {Serrano}, {Shafieloo}, {Shan}, {Sharples}, {Sholl},
  {Shourt}, {Silber}, {Silva}, {Sirk}, {Slosar}, {Smith}, {Smoot}, {Som},
  {Song}, {Sprayberry}, {Staten}, {Stefanik}, {Tarle}, {Sien Tie}, {Tinker},
  {Tojeiro}, {Valdes}, {Valenzuela}, {Valluri}, {Vargas-Magana}, {Verde},
  {Walker}, {Wang}, {Wang}, {Weaver}, {Weaverdyck}, {Wechsler}, {Weinberg},
  {White}, {Yang}, {Yeche}, {Zhang}, {Zhao}, {Zheng}, {Zhou}, {Zhou}, {Zhu},
  {Zou}, \& {Zu}}]{DESI1}
{DESI Collaboration} {et~al.}, 2016{\natexlab{a}}, arXiv e-prints,
  arXiv:1611.00036

\bibitem[{{DESI Collaboration} {et~al}\mbox{.}(2016{\natexlab{b}}){DESI
  Collaboration}, {Aghamousa}, {Aguilar}, {Ahlen}, {Alam}, {Allen}, {Allende
  Prieto}, {Annis}, {Bailey}, {Balland}, {Ballester}, {Baltay}, {Beaufore},
  {Bebek}, {Beers}, {Bell}, {Bernal}, {Besuner}, {Beutler}, {Blake}, {Bleuler},
  {Blomqvist}, {Blum}, {Bolton}, {Briceno}, {Brooks}, {Brownstein},
  {Buckley-Geer}, {Burden}, {Burtin}, {Busca}, {Cahn}, {Cai}, {Cardiel-Sas},
  {Carlberg}, {Carton}, {Casas}, {Castander}, {Cervantes-Cota}, {Claybaugh},
  {Close}, {Coker}, {Cole}, {Comparat}, {Cooper}, {Cousinou}, {Crocce}, {Cuby},
  {Cunningham}, {Davis}, {Dawson}, {de la Macorra}, {De Vicente}, {Delubac},
  {Derwent}, {Dey}, {Dhungana}, {Ding}, {Doel}, {Duan}, {Ealet}, {Edelstein},
  {Eftekharzadeh}, {Eisenstein}, {Elliott}, {Escoffier}, {Evatt}, {Fagrelius},
  {Fan}, {Fanning}, {Farahi}, {Farihi}, {Favole}, {Feng}, {Fernandez},
  {Findlay}, {Finkbeiner}, {Fitzpatrick}, {Flaugher}, {Flender}, {Font-Ribera},
  {Forero-Romero}, {Fosalba}, {Frenk}, {Fumagalli}, {Gaensicke}, {Gallo},
  {Garcia-Bellido}, {Gaztanaga}, {Pietro Gentile Fusillo}, {Gerard},
  {Gershkovich}, {Giannantonio}, {Gillet}, {Gonzalez-de-Rivera},
  {Gonzalez-Perez}, {Gott}, {Graur}, {Gutierrez}, {Guy}, {Habib}, {Heetderks},
  {Heetderks}, {Heitmann}, {Hellwing}, {Herrera}, {Ho}, {Holland}, {Honscheid},
  {Huff}, {Hutchinson}, {Huterer}, {Hwang}, {Illa Laguna}, {Ishikawa},
  {Jacobs}, {Jeffrey}, {Jelinsky}, {Jennings}, {Jiang}, {Jimenez}, {Johnson},
  {Joyce}, {Jullo}, {Juneau}, {Kama}, {Karcher}, {Karkar}, {Kehoe}, {Kennamer},
  {Kent}, {Kilbinger}, {Kim}, {Kirkby}, {Kisner}, {Kitanidis}, {Kneib},
  {Koposov}, {Kovacs}, {Koyama}, {Kremin}, {Kron}, {Kronig}, {Kueter-Young},
  {Lacey}, {Lafever}, {Lahav}, {Lambert}, {Lampton}, {Landriau}, {Lang},
  {Lauer}, {Le Goff}, {Le Guillou}, {Le Van Suu}, {Lee}, {Lee}, {Leitner},
  {Lesser}, {Levi}, {L'Huillier}, {Li}, {Liang}, {Lin}, {Linder}, {Loebman},
  {Luki{\'c}}, {Ma}, {MacCrann}, {Magneville}, {Makarem}, {Manera}, {Manser},
  {Marshall}, {Martini}, {Massey}, {Matheson}, {McCauley}, {McDonald},
  {McGreer}, {Meisner}, {Metcalfe}, {Miller}, {Miquel}, {Moustakas}, {Myers},
  {Naik}, {Newman}, {Nichol}, {Nicola}, {Nicolati da Costa}, {Nie}, {Niz},
  {Norberg}, {Nord}, {Norman}, {Nugent}, {O'Brien}, {Oh}, {Olsen}, {Padilla},
  {Padmanabhan}, {Padmanabhan}, {Palanque-Delabrouille}, {Palmese},
  {Pappalardo}, {P{\^a}ris}, {Park}, {Patej}, {Peacock}, {Peiris}, {Peng},
  {Percival}, {Perruchot}, {Pieri}, {Pogge}, {Pollack}, {Poppett}, {Prada},
  {Prakash}, {Probst}, {Rabinowitz}, {Raichoor}, {Ree}, {Refregier}, {Regal},
  {Reid}, {Reil}, {Rezaie}, {Rockosi}, {Roe}, {Ronayette}, {Roodman}, {Ross},
  {Ross}, {Rossi}, {Rozo}, {Ruhlmann-Kleider}, {Rykoff}, {Sabiu}, {Samushia},
  {Sanchez}, {Sanchez}, {Schlegel}, {Schneider}, {Schubnell}, {Secroun},
  {Seljak}, {Seo}, {Serrano}, {Shafieloo}, {Shan}, {Sharples}, {Sholl},
  {Shourt}, {Silber}, {Silva}, {Sirk}, {Slosar}, {Smith}, {Smoot}, {Som},
  {Song}, {Sprayberry}, {Staten}, {Stefanik}, {Tarle}, {Sien Tie}, {Tinker},
  {Tojeiro}, {Valdes}, {Valenzuela}, {Valluri}, {Vargas-Magana}, {Verde},
  {Walker}, {Wang}, {Wang}, {Weaver}, {Weaverdyck}, {Wechsler}, {Weinberg},
  {White}, {Yang}, {Yeche}, {Zhang}, {Zhao}, {Zheng}, {Zhou}, {Zhou}, {Zhu},
  {Zou}, \& {Zu}}]{DESI2}
{DESI Collaboration} {et~al.}, 2016{\natexlab{b}}, arXiv e-prints,
  arXiv:1611.00037

\bibitem[{{Donahue}, {Aldering} \& {Stocke}(1995){Donahue}, {Aldering}, \&
  {Stocke}}]{Donahue1995}
{Donahue} M., {Aldering} G., {Stocke} J.~T., 1995, \apjl, 450, L45

\bibitem[{{Donahue} \& {Voit}(2022)}]{Donahue2022}
{Donahue} M., {Voit} G.~M., 2022, \physrep, 973, 1

\bibitem[{{Dutta} {et~al}\mbox{.}(2020){Dutta}, {Fumagalli}, {Fossati},
  {Lofthouse}, {Prochaska}, {Arrigoni Battaia}, {Bielby}, {Cantalupo}, {Cooke},
  {Murphy}, \& {O'Meara}}]{Dutta2020}
{Dutta} R. {et~al.}, 2020, \mnras, 499, 5022

\bibitem[{{Fabian}(2012)}]{Fabian2012}
{Fabian} A.~C., 2012, \araa, 50, 455

\bibitem[{{Faucher-Gigu{\`e}re} \& {Oh}(2023)}]{CGM2023}
{Faucher-Gigu{\`e}re} C.-A., {Oh} S.~P., 2023, \araa, 61, 131

\bibitem[{{Fielding} {et~al}\mbox{.}(2020){Fielding}, {Tonnesen}, {DeFelippis},
  {Li}, {Su}, {Bryan}, {Kim}, {Forbes}, {Somerville}, {Battaglia}, {Schneider},
  {Li}, {Choi}, {Hayward}, \& {Hernquist}}]{Fielding2020}
{Fielding} D.~B. {et~al.}, 2020, \apj, 903, 32

\bibitem[{{Fumagalli} {et~al}\mbox{.}(2017){Fumagalli}, {Haardt}, {Theuns},
  {Morris}, {Cantalupo}, {Madau}, \& {Fossati}}]{Fumagalli2017}
{Fumagalli} M., {Haardt} F., {Theuns} T., {Morris} S.~L., {Cantalupo} S.,
  {Madau} P., {Fossati} M., 2017, \mnras, 467, 4802

\bibitem[{{Fumagalli} {et~al}\mbox{.}(2011){Fumagalli}, {Prochaska}, {Kasen},
  {Dekel}, {Ceverino}, \& {Primack}}]{Fumagalli2011}
{Fumagalli} M., {Prochaska} J.~X., {Kasen} D., {Dekel} A., {Ceverino} D.,
  {Primack} J.~R., 2011, \mnras, 418, 1796

\bibitem[{{Gallazzi} {et~al}\mbox{.}(2005){Gallazzi}, {Charlot}, {Brinchmann},
  {White}, \& {Tremonti}}]{Gallazzi}
{Gallazzi} A., {Charlot} S., {Brinchmann} J., {White} S.~D.~M., {Tremonti}
  C.~A., 2005, MNRAS, 362, 41

\bibitem[{{Haardt} \& {Madau}(1996)}]{Haardt1996}
{Haardt} F., {Madau} P., 1996, \apj, 461, 20

\bibitem[{{Haardt} \& {Madau}(2001)}]{Haardt2001}
{Haardt} F., {Madau} P., 2001, in Clusters of Galaxies and the High Redshift
  Universe Observed in X-rays, {Neumann} D.~M., {Tran} J.~T.~V., eds., p.~64

\bibitem[{{Haardt} \& {Madau}(2012)}]{Haardt2012}
{Haardt} F., {Madau} P., 2012, \apj, 746, 125

\bibitem[{{Herenz} {et~al}\mbox{.}(2023){Herenz}, {Inoue}, {Salas}, {Koenigs},
  {Moya-Sierralta}, {Cannon}, {Hayes}, {Papaderos}, {{\"O}stlin}, {Bik}, {Le
  Reste}, {Kusakabe}, {Monreal-Ibero}, \& {Puschnig}}]{Herenz2023}
{Herenz} E.~C. {et~al.}, 2023, \aap, 670, A121

\bibitem[{{Ho} {et~al}\mbox{.}(2017){Ho}, {Martin}, {Kacprzak}, \&
  {Churchill}}]{Ho2017}
{Ho} S.~H., {Martin} C.~L., {Kacprzak} G.~G., {Churchill} C.~W., 2017, \apj,
  835, 267

\bibitem[{{Hopkins}, {Quataert} \& {Murray}(2012){Hopkins}, {Quataert}, \&
  {Murray}}]{Hopkins2012}
{Hopkins} P.~F., {Quataert} E., {Murray} N., 2012, \mnras, 421, 3522

\bibitem[{{Hu}(2019)}]{Hu2019}
{Hu} C.-Y., 2019, \mnras, 483, 3363

\bibitem[{{Hummer} \& {Storey}(1987)}]{hummer}
{Hummer} D.~G., {Storey} P.~J., 1987, \mnras, 224, 801

\bibitem[{{Johnson}, {Chen} \& {Mulchaey}(2015){Johnson}, {Chen}, \&
  {Mulchaey}}]{Johnson2015}
{Johnson} S.~D., {Chen} H.-W., {Mulchaey} J.~S., 2015, \mnras, 449, 3263

\bibitem[{{Johnson} {et~al}\mbox{.}(2017){Johnson}, {Chen}, {Mulchaey},
  {Schaye}, \& {Straka}}]{Johnson2017}
{Johnson} S.~D., {Chen} H.-W., {Mulchaey} J.~S., {Schaye} J., {Straka} L.~A.,
  2017, \apjl, 850, L10

\bibitem[{{Johnson} {et~al}\mbox{.}(2014){Johnson}, {Chen}, {Mulchaey},
  {Tripp}, {Prochaska}, \& {Werk}}]{Johnson2014}
{Johnson} S.~D., {Chen} H.-W., {Mulchaey} J.~S., {Tripp} T.~M., {Prochaska}
  J.~X., {Werk} J.~K., 2014, \mnras, 438, 3039

\bibitem[{{Joshi} {et~al}\mbox{.}(2018){Joshi}, {Srianand}, {Petitjean}, \&
  {Noterdaeme}}]{joshi2018}
{Joshi} R., {Srianand} R., {Petitjean} P., {Noterdaeme} P., 2018, \mnras, 476,
  210

\bibitem[{{Kauffmann} {et~al}\mbox{.}(2003{\natexlab{a}}){Kauffmann},
  {Heckman}, {Tremonti}, {Brinchmann}, {Charlot}, {White}, {Ridgway}, \&
  et~al}]{kauffmann_agn}
{Kauffmann} G., {Heckman} T.~M., {Tremonti} C., {Brinchmann} J., {Charlot} S.,
  {White} S.~D.~M., {Ridgway} S.~E., et~al, 2003{\natexlab{a}}, \mnras, 346,
  1055

\bibitem[{{Kauffmann} {et~al}\mbox{.}(2003{\natexlab{b}}){Kauffmann},
  {Heckman}, {White}, {Charlot}, {Tremonti}, {Brinchmann}, {Bruzual}, {Peng},
  {Seibert}, {Bernardi}, {Blanton}, {Brinkmann}, {Castander}, {Cs{\'a}bai},
  {Fukugita}, {Ivezic}, {Munn}, {Nichol}, {Padmanabhan}, {Thakar}, {Weinberg},
  \& {York}}]{Kauffmann2003a}
{Kauffmann} G. {et~al.}, 2003{\natexlab{b}}, MNRAS, 341, 33

\bibitem[{{Kauffmann} {et~al}\mbox{.}(2003{\natexlab{c}}){Kauffmann},
  {Heckman}, {White}, {Charlot}, {Tremonti}, {Peng}, {Seibert}, \&
  et~al}]{Kauffmann2003b}
{Kauffmann} G., {Heckman} T.~M., {White} S.~D.~M., {Charlot} S., {Tremonti} C.,
  {Peng} E.~W., {Seibert} M., et~al, 2003{\natexlab{c}}, \mnras, 341, 54

\bibitem[{{Kewley} {et~al}\mbox{.}(2001){Kewley}, {Dopita}, {Sutherland},
  {Heisler}, \& {Trevena}}]{kewley}
{Kewley} L.~J., {Dopita} M.~A., {Sutherland} R.~S., {Heisler} C.~A., {Trevena}
  J., 2001, \apj, 556, 121

\bibitem[{{Khabibullin} {et~al}\mbox{.}(2023){Khabibullin}, {Galeazzi},
  {Bogdan}, {Cann}, {Churazov}, {Dolag}, {Drake}, {Forman}, {Hernquist},
  {Koutroumpa}, {Kraft}, {Kuntz}, {Markevitch}, {McCammon}, {Ogorzalek},
  {Pfeifle}, {Pillepich}, {Plucinsky}, {Ponti}, {Schellenberger}, {Truong},
  {Valentini}, {Veilleux}, {Vladutescu-Zopp}, {Wang}, \&
  {Weaver}}]{Khabibullin2023}
{Khabibullin} I. {et~al.}, 2023, arXiv e-prints, arXiv:2310.16038

\bibitem[{{Kim} \& {Ostriker}(2018)}]{Kim2018}
{Kim} C.-G., {Ostriker} E.~C., 2018, \apj, 853, 173

\bibitem[{{Kraft} {et~al}\mbox{.}(2022){Kraft}, {Markevitch}, {Kilbourne},
  {Adams}, {Akamatsu}, {Ayromlou}, {Bandler}, {Barbera}, {Bennett}, {Bhardwaj},
  {Biffi}, {Bodewits}, {Bogdan}, {Bonamente}, {Borgani}, {Branduardi-Raymont},
  {Bregman}, {Burchett}, {Cann}, {Carter}, {Chakraborty}, {Churazov}, {Crain},
  {Cumbee}, {Dave}, {DiPirro}, {Dolag}, {Bertrand Doriese}, {Drake}, {Dunn},
  {Eckart}, {Eckert}, {Ettori}, {Forman}, {Galeazzi}, {Gall}, {Gatuzz}, {Hell},
  {Hodges-Kluck}, {Jackman}, {Jahromi}, {Jennings}, {Jones}, {Kaaret},
  {Kavanagh}, {Kelley}, {Khabibullin}, {Kim}, {Koutroumpa}, {Kovacs}, {Kuntz},
  {Lau}, {Lee}, {Leutenegger}, {Lin}, {Lisse}, {Lo Cicero}, {Lovisari},
  {McCammon}, {McEntee}, {Mernier}, {Miller}, {Nagai}, {Negro}, {Nelson},
  {Ness}, {Nulsen}, {Ogorzalek}, {Oppenheimer}, {Oskinova}, {Patnaude},
  {Pfeifle}, {Pillepich}, {Plucinsky}, {Pooley}, {Porter}, {Randall}, {Rasia},
  {Raymond}, {Ruszkowski}, {Sakai}, {Sarkar}, {Sasaki}, {Sato},
  {Schellenberger}, {Schaye}, {Simionescu}, {Smith}, {Steiner}, {Stern}, {Su},
  {Sun}, {Tremblay}, {Truong}, {Tutt}, {Ursino}, {Veilleux}, {Vikhlinin},
  {Vladutescu-Zopp}, {Vogelsberger}, {Walker}, {Weaver}, {Weigt}, {Werk},
  {Werner}, {Wolk}, {Zhang}, {Zhang}, {Zhuravleva}, \& {ZuHone}}]{Kraft2022}
{Kraft} R. {et~al.}, 2022, arXiv e-prints, arXiv:2211.09827

\bibitem[{{Lan} \& {Mo}(2018)}]{lan2018}
{Lan} T.-W., {Mo} H., 2018, ArXiv 1806.05786

\bibitem[{{Levesque}, {Kewley} \& {Larson}(2010){Levesque}, {Kewley}, \&
  {Larson}}]{Levesque2010}
{Levesque} E.~M., {Kewley} L.~J., {Larson} K.~L., 2010, \aj, 139, 712

\bibitem[{{Li} \& {Bryan}(2020)}]{Limiao2020}
{Li} M., {Bryan} G.~L., 2020, \apjl, 890, L30

\bibitem[{{Li}, {Bryan} \& {Ostriker}(2017){Li}, {Bryan}, \&
  {Ostriker}}]{Li2017}
{Li} M., {Bryan} G.~L., {Ostriker} J.~P., 2017, \apj, 841, 101

\bibitem[{{Li} \& {Tonnesen}(2020)}]{Li2020}
{Li} M., {Tonnesen} S., 2020, \apj, 898, 148

\bibitem[{{Martin} {et~al}\mbox{.}(2019){Martin}, {Ho}, {Kacprzak}, \&
  {Churchill}}]{Martin2019}
{Martin} C.~L., {Ho} S.~H., {Kacprzak} G.~G., {Churchill} C.~W., 2019, \apj,
  878, 84

\bibitem[{{Mart{\'\i}n-Navarro} {et~al}\mbox{.}(2021){Mart{\'\i}n-Navarro},
  {Pillepich}, {Nelson}, {Rodriguez-Gomez}, {Donnari}, {Hernquist}, \&
  {Springel}}]{Ignacio2021}
{Mart{\'\i}n-Navarro} I., {Pillepich} A., {Nelson} D., {Rodriguez-Gomez} V.,
  {Donnari} M., {Hernquist} L., {Springel} V., 2021, \nat, 594, 187

\bibitem[{{Mitchell} {et~al}\mbox{.}(2020){Mitchell}, {Schaye}, {Bower}, \&
  {Crain}}]{Mitchell2020}
{Mitchell} P.~D., {Schaye} J., {Bower} R.~G., {Crain} R.~A., 2020, \mnras, 494,
  3971

\bibitem[{{Morganti}(2017)}]{Morganti2017}
{Morganti} R., 2017, Frontiers in Astronomy and Space Sciences, 4, 42

\bibitem[{{Naab} \& {Ostriker}(2017)}]{Naab2017}
{Naab} T., {Ostriker} J.~P., 2017, \araa, 55, 59

\bibitem[{{Nelson} {et~al}\mbox{.}(2019){Nelson}, {Pillepich}, {Springel},
  {Pakmor}, {Weinberger}, {Genel}, {Torrey}, {Vogelsberger}, {Marinacci}, \&
  {Hernquist}}]{Nelson2019}
{Nelson} D. {et~al.}, 2019, \mnras, 490, 3234

\bibitem[{{Nielsen} {et~al}\mbox{.}(2023){Nielsen}, {Fisher}, {Kacprzak},
  {Chisholm}, {Martin}, {Reichardt Chu}, {Sandstrom}, \& {Rickards
  Vaught}}]{Nielsen2023}
{Nielsen} N.~M., {Fisher} D.~B., {Kacprzak} G.~G., {Chisholm} J., {Martin}
  D.~C., {Reichardt Chu} B., {Sandstrom} K.~M., {Rickards Vaught} R.~J., 2023,
  arXiv e-prints, arXiv:2311.00856

\bibitem[{{Norris} {et~al}\mbox{.}(2021){Norris}, {Muzahid}, {Charlton},
  {Kacprzak}, {Wakker}, \& {Churchill}}]{Norris2021}
{Norris} J.~M., {Muzahid} S., {Charlton} J.~C., {Kacprzak} G.~G., {Wakker}
  B.~P., {Churchill} C.~W., 2021, \mnras, 506, 5640

\bibitem[{{Osterbrock} \& {Ferland}(2006)}]{osterbrock2006}
{Osterbrock} D.~E., {Ferland} G.~J., 2006, {Astrophysics of gaseous nebulae and
  active galactic nuclei}

\bibitem[{{P{\'e}roux} {et~al}\mbox{.}(2020){P{\'e}roux}, {Nelson}, {van de
  Voort}, {Pillepich}, {Marinacci}, {Vogelsberger}, \&
  {Hernquist}}]{Peroux2020}
{P{\'e}roux} C., {Nelson} D., {van de Voort} F., {Pillepich} A., {Marinacci}
  F., {Vogelsberger} M., {Hernquist} L., 2020, \mnras, 499, 2462

\bibitem[{{Pillepich} {et~al}\mbox{.}(2019){Pillepich}, {Nelson}, {Springel},
  {Pakmor}, {Torrey}, {Weinberger}, {Vogelsberger}, {Marinacci}, {Genel}, {van
  der Wel}, \& {Hernquist}}]{Pillepich2019}
{Pillepich} A. {et~al.}, 2019, \mnras, 490, 3196

\bibitem[{{Planck Collaboration} {et~al}\mbox{.}(2018){Planck Collaboration},
  {Akrami}, {Arroja}, {Ashdown}, {Aumont}, {Baccigalupi}, {Ballardini},
  {Banday}, {Barreiro}, \& {Bartolo}}]{Planck2018}
{Planck Collaboration} {et~al.}, 2018, arXiv e-prints, arXiv:1807.06205

\bibitem[{{Pointon} {et~al}\mbox{.}(2019){Pointon}, {Kacprzak}, {Nielsen},
  {Muzahid}, {Murphy}, {Churchill}, \& {Charlton}}]{Pointon2019}
{Pointon} S.~K., {Kacprzak} G.~G., {Nielsen} N.~M., {Muzahid} S., {Murphy}
  M.~T., {Churchill} C.~W., {Charlton} J.~C., 2019, \apj, 883, 78

\bibitem[{{Prochaska} {et~al}\mbox{.}(2017){Prochaska}, {Werk}, {Worseck},
  {Tripp}, {Tumlinson}, \& et~al}]{prochaska2017}
{Prochaska} J.~X., {Werk} J.~K., {Worseck} G., {Tripp} T.~M., {Tumlinson} J.,
  et~al, 2017, \apj, 837, 169

\bibitem[{{Qu} \& {Bregman}(2022)}]{Qu2022}
{Qu} Z., {Bregman} J.~N., 2022, \apj, 927, 228

\bibitem[{{Ramesh} {et~al}\mbox{.}(2023){Ramesh}, {Nelson}, {Heesen}, \&
  {Br{\"u}ggen}}]{Ramesh2023a}
{Ramesh} R., {Nelson} D., {Heesen} V., {Br{\"u}ggen} M., 2023, arXiv e-prints,
  arXiv:2305.11214

\bibitem[{{Riess} {et~al}\mbox{.}(2018){Riess}, {Casertano}, {Yuan}, {Macri},
  {Bucciarelli}, {Lattanzi}, {MacKenty}, {Bowers}, {Zheng}, {Filippenko},
  {Huang}, \& {Anderson}}]{riess}
{Riess} A.~G. {et~al.}, 2018, \apj, 861, 126

\bibitem[{{Schaye} {et~al}\mbox{.}(2015){Schaye}, {Crain}, {Bower}, {Furlong},
  {Schaller}, {Theuns}, {Dalla Vecchia}, {Frenk}, {McCarthy}, {Helly},
  {Jenkins}, {Rosas-Guevara}, {White}, {Baes}, {Booth}, {Camps}, {Navarro},
  {Qu}, {Rahmati}, {Sawala}, {Thomas}, \& {Trayford}}]{EAGLE2}
{Schaye} J. {et~al.}, 2015, \mnras, 446, 521

\bibitem[{{Schroetter} {et~al}\mbox{.}(2019){Schroetter}, {Bouch{\'e}}, {Zabl},
  {Contini}, {Wendt}, {Schaye}, {Mitchell}, {Muzahid}, {Marino}, {Bacon},
  {Lilly}, {Richard}, \& {Wisotzki}}]{Schroetter2019}
{Schroetter} I. {et~al.}, 2019, \mnras, 490, 4368

\bibitem[{{Shen} {et~al}\mbox{.}(2013){Shen}, {Madau}, {Guedes}, {Mayer},
  {Prochaska}, \& {Wadsley}}]{Shen2013}
{Shen} S., {Madau} P., {Guedes} J., {Mayer} L., {Prochaska} J.~X., {Wadsley}
  J., 2013, \apj, 765, 89

\bibitem[{{Simard} {et~al}\mbox{.}(2011){Simard}, {Mendel}, {Patton},
  {Ellison}, \& {McConnachie}}]{simard}
{Simard} L., {Mendel} J.~T., {Patton} D.~R., {Ellison} S.~L., {McConnachie}
  A.~W., 2011, ApJs, 196, 11

\bibitem[{{Somerville} \& {Dav{\'e}}(2015)}]{Somerville2015}
{Somerville} R.~S., {Dav{\'e}} R., 2015, \araa, 53, 51

\bibitem[{{Steidel} {et~al}\mbox{.}(2010){Steidel}, {Erb}, {Shapley},
  {Pettini}, {Reddy}, {Bogosavljevi{\'c}}, {Rudie}, \& {Rakic}}]{steidel2010}
{Steidel} C.~C., {Erb} D.~K., {Shapley} A.~E., {Pettini} M., {Reddy} N.,
  {Bogosavljevi{\'c}} M., {Rudie} G.~C., {Rakic} O., 2010, \apj, 717, 289

\bibitem[{{Tremonti} {et~al}\mbox{.}(2004){Tremonti}, {Heckman}, {Kauffmann},
  {Brinchmann}, {Charlot}, {White}, {Seibert}, {Peng}, {Schlegel}, {Uomoto},
  {Fukugita}, \& {Brinkmann}}]{Tremonti2004}
{Tremonti} C.~A. {et~al.}, 2004, \apj, 613, 898

\bibitem[{{Truong} {et~al}\mbox{.}(2021){Truong}, {Pillepich}, {Nelson},
  {Werner}, \& {Hernquist}}]{Truong2021}
{Truong} N., {Pillepich} A., {Nelson} D., {Werner} N., {Hernquist} L., 2021,
  \mnras, 508, 1563

\bibitem[{{Tumlinson}, {Peeples} \& {Werk}(2017){Tumlinson}, {Peeples}, \&
  {Werk}}]{CGM2017}
{Tumlinson} J., {Peeples} M.~S., {Werk} J.~K., 2017, \araa, 55, 389

\bibitem[{{Veilleux} \& {Osterbrock}(1987)}]{vo}
{Veilleux} S., {Osterbrock} D.~E., 1987, \apjs, 63, 295

\bibitem[{{Werk} {et~al}\mbox{.}(2016){Werk}, {Prochaska}, {Cantalupo}, {Fox},
  {Oppenheimer}, {Tumlinson}, {Tripp}, \& et~al}]{werk16}
{Werk} J.~K., {Prochaska} J.~X., {Cantalupo} S., {Fox} A.~J., {Oppenheimer} B.,
  {Tumlinson} J., {Tripp} T.~M., et~al, 2016, \apj, 833, 54

\bibitem[{{Werk} {et~al}\mbox{.}(2013){Werk}, {Prochaska}, {Thom}, {Tumlinson},
  {Tripp}, {O'Meara}, \& {Peeples}}]{Werk2013}
{Werk} J.~K., {Prochaska} J.~X., {Thom} C., {Tumlinson} J., {Tripp} T.~M.,
  {O'Meara} J.~M., {Peeples} M.~S., 2013, \apjs, 204, 17

\bibitem[{{Werk} {et~al}\mbox{.}(2014){Werk}, {Prochaska}, {Tumlinson},
  {Peeples}, {Tripp}, {Fox}, \& et~al}]{Werk2014}
{Werk} J.~K., {Prochaska} J.~X., {Tumlinson} J., {Peeples} M.~S., {Tripp}
  T.~M., {Fox} A.~J., et~al, 2014, \apj, 792, 8

\bibitem[{{Wilde} {et~al}\mbox{.}(2021){Wilde}, {Werk}, {Burchett},
  {Prochaska}, {Tchernyshyov}, {Tripp}, {Tejos}, {Lehner}, {Bordoloi},
  {O'Meara}, \& {Tumlinson}}]{CGMsquare2021}
{Wilde} M.~C. {et~al.}, 2021, \apj, 912, 9

\bibitem[{{York} {et~al}\mbox{.}(2000){York}, {Adelman}, {Anderson},
  {Anderson}, {Annis}, {Bahcall}, {Bakken}, {Barkhouser}, {Bastian}, {Berman},
  {Boroski}, {Bracker}, {Briegel}, {Briggs}, {Brinkmann}, {Brunner}, {Burles},
  {Carey}, {Carr}, {Castander}, {Chen}, {Colestock}, {Connolly}, {Crocker},
  {Csabai}, {Czarapata}, {Davis}, {Doi}, {Dombeck}, {Eisenstein}, {Ellman},
  {Elms}, {Evans}, {Fan}, {Federwitz}, {Fiscelli}, {Friedman}, {Frieman},
  {Fukugita}, {Gillespie}, {Gunn}, {Gurbani}, {de Haas}, {Haldeman}, {Harris},
  {Hayes}, {Heckman}, {Hennessy}, {Hindsley}, {Holm}, {Holmgren}, {Huang},
  {Hull}, {Husby}, {Ichikawa}, {Ichikawa}, {Ivezi{\'c}}, {Kent}, {Kim},
  {Kinney}, {Klaene}, {Kleinman}, {Kleinman}, {Knapp}, {Korienek}, {Kron},
  {Kunszt}, {Lamb}, {Lee}, {Leger}, {Limmongkol}, {Lindenmeyer}, {Long},
  {Loomis}, {Loveday}, {Lucinio}, {Lupton}, {MacKinnon}, {Mannery}, {Mantsch},
  {Margon}, {McGehee}, {McKay}, {Meiksin}, {Merelli}, {Monet}, {Munn},
  {Narayanan}, {Nash}, {Neilsen}, {Neswold}, {Newberg}, {Nichol}, {Nicinski},
  {Nonino}, {Okada}, {Okamura}, {Ostriker}, {Owen}, {Pauls}, {Peoples},
  {Peterson}, {Petravick}, {Pier}, {Pope}, {Pordes}, {Prosapio},
  {Rechenmacher}, {Quinn}, {Richards}, {Richmond}, {Rivetta}, {Rockosi},
  {Ruthmansdorfer}, {Sandford}, {Schlegel}, {Schneider}, {Sekiguchi}, {Sergey},
  {Shimasaku}, {Siegmund}, {Smee}, {Smith}, {Snedden}, {Stone}, {Stoughton},
  {Strauss}, {Stubbs}, {SubbaRao}, {Szalay}, {Szapudi}, {Szokoly}, {Thakar},
  {Tremonti}, {Tucker}, {Uomoto}, {Vanden Berk}, {Vogeley}, {Waddell}, {Wang},
  {Watanabe}, {Weinberg}, {Yanny}, {Yasuda}, \& {SDSS
  Collaboration}}]{SDSS2000}
{York} D.~G. {et~al.}, 2000, \aj, 120, 1579

\bibitem[{{Yoshida} {et~al}\mbox{.}(2016){Yoshida}, {Yagi}, {Ohyama},
  {Komiyama}, {Kashikawa}, {Tanaka}, \& {Okamura}}]{Yoshida2016}
{Yoshida} M., {Yagi} M., {Ohyama} Y., {Komiyama} Y., {Kashikawa} N., {Tanaka}
  H., {Okamura} S., 2016, \apj, 820, 48

\bibitem[{{Zahedy} {et~al}\mbox{.}(2019){Zahedy}, {Chen}, {Johnson}, {Pierce},
  {Rauch}, {Huang}, {Weiner}, \& {Gauthier}}]{Zahedy2019}
{Zahedy} F.~S., {Chen} H.-W., {Johnson} S.~D., {Pierce} R.~M., {Rauch} M.,
  {Huang} Y.-H., {Weiner} B.~J., {Gauthier} J.-R., 2019, \mnras, 484, 2257

\bibitem[{{Zhang} {et~al}\mbox{.}(2020{\natexlab{a}}){Zhang}, {Fang},
  {Zaritsky}, {Behroozi}, {Werk}, \& {Yang}}]{Zhang2020b}
{Zhang} H., {Fang} T., {Zaritsky} D., {Behroozi} P., {Werk} J., {Yang} X.,
  2020{\natexlab{a}}, \apjl, 893, L3 (Paper VI)

\bibitem[{{Zhang} {et~al}\mbox{.}(2020{\natexlab{b}}){Zhang}, {Yang},
  {Zaritsky}, {Behroozi}, \& {Werk}}]{Zhang2020a}
{Zhang} H., {Yang} X., {Zaritsky} D., {Behroozi} P., {Werk} J.,
  2020{\natexlab{b}}, \apj, 880, 33 (Paper V)

\bibitem[{{Zhang} \& {Zaritsky}(2022)}]{Zhang2022}
{Zhang} H., {Zaritsky} D., 2022, \apj, 941, 18 (Paper VIII)

\bibitem[{{Zhang}, {Zaritsky} \& {Behroozi}(2018){Zhang}, {Zaritsky}, \&
  {Behroozi}}]{zhang2018a}
{Zhang} H., {Zaritsky} D., {Behroozi} P., 2018, \apj, 861, 34 (Paper II)

\bibitem[{{Zhang} {et~al}\mbox{.}(2019){Zhang}, {Zaritsky}, {Behroozi}, \&
  {Werk}}]{Zhang2019}
{Zhang} H., {Zaritsky} D., {Behroozi} P., {Werk} J., 2019, \apj, 880, 28 (Paper
  IV)

\bibitem[{{Zhang} {et~al}\mbox{.}(2021){Zhang}, {Zaritsky}, {Olsen},
  {Behroozi}, {Werk}, {Kennicutt}, {Xie}, {Yang}, {Fang}, {De Lucia},
  {Hirschmann}, \& {Fontanot}}]{Zhang2021}
{Zhang} H. {et~al.}, 2021, \apj, 916, 101 (Paper VII)

\bibitem[{{Zhang} {et~al}\mbox{.}(2018){Zhang}, {Zaritsky}, {Werk}, \&
  {Behroozi}}]{Zhang2018b}
{Zhang} H., {Zaritsky} D., {Werk} J., {Behroozi} P., 2018, \apj, 866, L4 (Paper
  III)

\bibitem[{{Zhang} {et~al}\mbox{.}(2016){Zhang}, {Zaritsky}, {Zhu},
  {M{\'e}nard}, \& {Hogg}}]{zhang2016}
{Zhang} H., {Zaritsky} D., {Zhu} G., {M{\'e}nard} B., {Hogg} D.~W., 2016, \apj,
  833, 276 (Paper I)

\bibitem[{{Zhu} \& {M{\'e}nard}(2013{\natexlab{a}})}]{zhu2013a}
{Zhu} G., {M{\'e}nard} B., 2013{\natexlab{a}}, \apj, 773, 16

\bibitem[{{Zhu} \& {M{\'e}nard}(2013{\natexlab{b}})}]{zhu2013b}
{Zhu} G., {M{\'e}nard} B., 2013{\natexlab{b}}, \apj, 770, 130

\end{thebibliography}

\end{document}